\documentclass
[12pt,
aps,
pra,
preprint,
tightenlines,
amsmath,
amssymb,
superscriptaddress,
onecolumn,
notitlepage, 
nofootinbib]{revtex4-2}
\usepackage{amssymb, amsbsy, amsmath, latexsym, dsfont, array, layout, graphicx, mathrsfs, color,soul}
\usepackage{xcolor}
\definecolor{darkred}{rgb}{0.5,0,0}
\definecolor{darkgreen}{rgb}{0,0.5,0}
\definecolor{darkblue}{rgb}{0,0,0.5}

\usepackage[pdfpagemode=UseOutlines,
  colorlinks=true, bookmarks=true, bookmarksnumbered=true]{hyperref} % pagebackref=true
\hypersetup{colorlinks, linkcolor=darkblue, filecolor=darkgreen, urlcolor=darkblue, citecolor=darkblue}

\usepackage{siunitx}
\usepackage{enumitem}

\newcommand{\ket}[1]{\left|{#1}\right\rangle}
\newcommand{\bra}[1]{\left\langle{#1}\right|}

\newcommand{\expect}[1]{\langle{#1}\rangle}

\newcommand{\bnm}{\begin{enumerate}}
\newcommand{\edm}{\end{enumerate}}

\usepackage[normalem]{ulem}%This enables strike through text using the \sout{} command.
\usepackage{bbold}
\usepackage{comment}

\NewDocumentEnvironment{myprotocol}{m +b}{%
  \parindent0em 
  \begin{minipage}{\linewidth}
    \hrule width \hsize height 1pt \kern 1mm \hrule width \hsize 
    \vspace{0.1cm}
    \begin{center} 
      \textbf{#1}
    \end{center}
  \end{minipage}
    \hrule width \hsize \kern 1mm \hrule width \hsize height 1pt 
    \vspace{5pt}
    #2
    \noindent\hrulefill
  }{}

\begin{document}

\title{The EU Quantum Flagship's Key Performance Indicators for Quantum Computing}

%\author{Prepared by Benchmark Teams of OpenSuperQPlus}
%\affiliation{https://opensuperqplus.eu/}
\author{Zolt\'{a}n Zimbor\'{a}s}
\affiliation{HUN-REN Wigner Research Centre for Physics, 1525 P.O. Box 49, Hungary}
\affiliation{Department of Physics, University of Helsinki, Helsinki, Finland}
\affiliation{Algorithmiq Ltd., Helsinki, Finland}
\author{Attila Portik}
\affiliation{HUN-REN Wigner Research Centre for Physics, 1525 P.O. Box 49, Hungary}
\affiliation{E\"otv\"os Lor\'and University, P.O.\ Box 32, H-1518, Hungary}
\affiliation{Qutility @ Faulhorn Labs,  H-1117 Budapest, Hungary}
\author{David Aguirre}
\affiliation{Department of Physical Chemistry, University of the Basque Country UPV/EHU,  48080 Bilbao, Spain}
\affiliation{BCAM - Basque Center for Applied Mathematics, 48009 Bilbao, Spain}
\author{Rub\'{e}n Pe\~{n}a}
\affiliation{BCAM - Basque Center for Applied Mathematics, 48009 Bilbao, Spain}
\author{Domonkos Svastits}
\affiliation{Department of Theoretical Physics, Institute of Physics, Budapest University of Technology and Economics, H-1111 Budapest, Hungary}
\affiliation{Qutility @ Faulhorn Labs, H-1117 Budapest, Hungary}
\author{Andr\'{a}s P\'{a}lyi}
\affiliation{Department of Theoretical Physics, Institute of Physics, Budapest University of Technology and Economics, H-1111 Budapest, Hungary}
\affiliation{HUN-REN-BME-BCE Quantum Technology Research Group, H-1111 Budapest, Hungary}
\author{Áron Márton}
\affiliation{Forschungszenctrum Jülich GmbH, 52428 Jülich, Germany}
\affiliation{ RWTH Aachen University, 52056 Aachen, Germany}
\author{J\'{a}nos K. Asb\'{o}th}
\affiliation{Department of Theoretical Physics, Institute of Physics, Budapest University of Technology and Economics, H-1111 Budapest, Hungary}
\affiliation{HUN-REN-BME-BCE Quantum Technology Research Group, H-1111 Budapest, Hungary}
\affiliation{HUN-REN Wigner Research Centre for Physics, 1525 P.O. Box 49, Hungary}
\author{Anton Frisk Kockum}
\affiliation{Department of Microtechnology and Nanoscience, Chalmers University of Technology, 412 96 Gothenburg, Sweden}
\author{Mikel Sanz}
  \affiliation{Department of Physical Chemistry, University of the Basque
  Country UPV/EHU,  48080 Bilbao, Spain}
  \affiliation{BCAM - Basque Center for Applied Mathematics, 48009 Bilbao, Spain}
  \affiliation{EHU Quantum Center, University of the Basque Country UPV/EHU, 48080 Bilbao, Spain}
  \affiliation{IKERBASQUE, Basque Foundation for Science,  48009, Bilbao, Spain}
\author{Orsolya K\'{a}lm\'{a}n}
\affiliation{HUN-REN Wigner Research Centre for Physics, 1525 P.O. Box 49, Hungary}
\author{Thomas Monz}
\affiliation{University of Innsbruck, Institute of Experimental Physics, 6020 Innsbruck, Austria}
\affiliation{Alpine Quantum Technologies GmbH, 6020 Innsbruck, Austria}
\author{Frank Wilhelm-Mauch}
\affiliation{Forschungszenctrum Jülich GmbH, 52428 Jülich, Germany}
\affiliation{Theoretical Physics, Universität des Saarlandes, 66123 Saarbrücken, Germany}

%\footnote{Authors: A. Portik, O. Kálmán, Z. Zimborás, A. Pályi, J. Asbóth \\ [-1mm] 
%Contributors: A. Frisk Kockum, R. Peña,  A. Portik, M. Sanz, F. Wilhelm-Mauch} 

%\author{Millenion}
%\affiliation{https://www.millenion.eu/ \\ \vspace{6ex}}
%\footnote{Contributors: A. Erhard, C. Marciniak, T. Monz}  \\

%Author 1$^{1}$, Author 2$^{2,1}$, Author 3$^{2}$, Author 4$^{1}$ \\

%\small $^{1}$\textit{Institute 1} \\
%$^{2}$\textit{Institute 2}\\

\begin{abstract}
\vspace{1.5ex}
As quantum processors continue to scale in size and complexity, the need for well-defined, reproducible, and technology-agnostic performance metrics becomes increasingly critical. Here we present a suite of scalable quantum computing benchmarks developed as key performance indicators (KPIs) within the EU Quantum Flagship. These proposed benchmarks are designed to assess holistic system performance rather than isolated components, and to remain applicable across both noisy intermediate-scale quantum (NISQ) devices and future fault-tolerant architectures. We introduce four core benchmarks addressing complementary aspects of quantum computing capability: large multi-qubit circuit execution via a Clifford Volume benchmark, scalable multipartite entanglement generation through GHZ-state preparation, a benchmark based on the application of Shor’s period-finding subroutine to simple functions, and a protocol quantifying the benefit of quantum error correction using Bell states. Each benchmark is accompanied by clearly specified protocols, reporting standards, and scalable evaluation methods. Together, these KPIs provide a coherent framework for transparent and fair performance assessment across quantum hardware platforms and for tracking progress late-NISQ toward early fault-tolerant quantum computation.
\end{abstract}

\maketitle
\newpage

\tableofcontents
\newpage

\section{Introduction}

%As quantum computing advances, evaluating the performance of quantum processors has become an essential task
Properly assessing the performance of quantum processors is essential to ensure that progress in quantum computing is guided by meaningful and reliable benchmarks~\cite{eisert2020quantum, proctor2024benchmarking,
Hashim2025TutorialQCVV,David,lorenz2025systematic,lall2025review}. 
%Given the rapid development, scalability emerges as a characteristic that benchmarks should satisfy when evaluating quantum devices. 
Given the rapid technological advancements and the development of new quantum processors with larger numbers of qubits (hundreds and more)~\cite{Acharya2025, Bluvstein2025, Gao2025, Photonic, Ransford2025, Kim2023}, scalability emerges as a characteristic that benchmarks must satisfy in order to effectively evaluate quantum computers.
Many existing benchmark procedures suffer from scalability
issues~\cite{PhysRevA.100.032328, eisert2020quantum, proctor2024benchmarking}, making them unsuitable for characterizing quantum processors with more than $\sim$50 qubits. 
Furthermore, considering the different technologies available for implementing quantum computing today~\cite{Acharya2025, Bluvstein2025, Gao2025, Photonic, Ransford2025, Kim2023, Chen2024benchmarkingtrapped,IQMs,Henriet2020quantumcomputing,Photonic, Edlbauer2025}, it is essential that benchmarks remain platform-agnostic, enabling fair and unbiased comparisons across different platforms.
Additionally, benchmarks should be well-defined, with clearly established protocols and rules that leave no room for ambiguity. This ensures that the benchmark can be consistently reproduced, allowing researchers and practitioners to follow the same methodology, leading to reliable and reproducible results. 
A fair, reproducible, and well-defined benchmark guarantees that the evaluation of quantum computer performance will be transparent and impartial, which is essential for the progress of quantum computing. 

Taking these considerations into account,  we have developed benchmarks for the Quantum Flagship program of the European Union as a suite of key performance indicators (KPIs).
Quantum computer benchmarks, to date, have mainly been protocols that were developed by the quantum community. Not all of them follow as rigorous and as well-established standards as benchmarks \cite{proctor2024benchmarking,
Hashim2025TutorialQCVV,David,lorenz2025systematic}. In the following, we therefore aim to provide a best-practices library of quantum computing KPIs that (a) explains the benchmark protocols, (b) provides standardised evaluation routines taking into account statistical aspects,
(c) contains a software development kit that implements the protocols in a hardware-agnostic way, and (d) provides sample data that also allows one to improve the code using continuous integration to check the validity of code changes. 
%We have deliberately excluded fully algorithmic benchmarks, which are most meaningful for fully fault-tolerant quantum computers. In the late-NISQ and early fault-tolerant eras, practically useful algorithms will likely remain sporadic and platform-dependent. Therefore, while monitoring progress remains important, as demonstrated by the recently launched Quantum Advantage Tracker, a different strategy is needed to quantify computational capacity in ways that enable direct comparisons.

The proposed benchmarks are detailed in Secs.~\ref{sec:clifford-volume-benchmark}--\ref{sec:qec-benchmark} of this document. In addition, the protocols are made publicly available on an open repository using OpenQASM, such that they are readily implementable on existing hardware~\cite{EQCB}. Beside, providing the protocol implementation, we also provide tools for the statistical evaluation. In addition to the protocol implementation, we provide tools for statistical evaluation. Sample data from partners are collected and shared to enable community-driven improvements to the benchmarking suite.

\section{Overview of the benchmarks}

We present four core KPI benchmarks, covering a wide range of aspects: from holistic system benchmarks, via demonstrations of scalability towards fault-tolerant performance, to algorithmic primitives. One important motivation for this breadth is to avoid that the quantum-computing community focuses on improving just a single aspect of their systems.

%The chosen KPIs are scalable, cover holistic system descriptions, include quantum applications, and are suitable to demonstrate capabilities both in the NISQ and FTQC era:

%We based our selection address aspects of the performance of a quantum computer \textit{system}, not a single component

%As engineered quantum systems evolve and mature, the number of qubits will increase. The benchmarks should thus be \emph{scalable} and, in particular, should also be suitable in a domain where classical computers can no longer mimic a quantum computer. We therefore constrain our selection to benchmarks that can be implemented \emph{both in the current noisy-intermediate scale quantum (NISQ) domain}~\cite{preskill2018quantum} as well as in the \emph{fault-tolerant quantum computing (FTQC) domain}~~\cite{gottesman1998theory}. 
%The goal here is to maintain the possibility to communicate progress in the community. As the field is continuously evolving, we foresee that these benchmarks shall undergo a similar evolution. We also anticipate that modifications will be initiated based on the feedback from the community.

\emph{Deep multi-qubit quantum circuits:} The Quantum Volume (QV) benchmark~\cite{PhysRevA.100.032328} is well established; however, it requires classical evaluation of the so-called heavy outputs, which is not a scalable procedure. Thus, we propose to apply an alternative protocol, which we refer to as the Clifford Volume (CLV) benchmark~\cite{CliffordVolume2025}. Focusing on Clifford gate operations, the output state in the CLV benchmark can be characterized in a scalable way. Moreover, since random Clifford operations are defined abstractly, without referring to a particular gate decomposition structure, unlike in the case of the QV, it can be regarded as more platform-independent than the QV. Still, the CLV benchmark maintains the volumetric aspect of QV, covering both the number of qubits in the register as well as the depth of the computation. Another advantage of the CLV benchmark is that Clifford gates can also be readily implemented for (most) logical encodings, making this protocol suitable also for the early fault tolerant domain. The details of the protocol are given in Sec.~\ref{sec:clifford-volume-benchmark}.

\emph{Multi-partite entanglement:} The task of preparing Greenberger--Horne--Zeilinger (GHZ) states~\cite{Greenberger1990} is a historically well established system benchmark~\cite{Monz2011, Pogorelov2021}, %, offer genuine multipartite entanglement, and 
which can be readily evaluated in a scalable fashion, e.g., using stabiliser measurements. The creation and evaluation of the GHZ state requires local and entangling gates across the quantum register, thus offering a volumetric benchmark. The underlying structure of the state is, furthermore, readily extendable towards logical qubits, as has been demonstrated in Ref.~\cite{Bluvstein2023}. For details of the protocol and its evaluation, see Sec.~\ref{sec:ghz-benchmark}.

\emph{Shor's period-finding algorithm:} The factoring algorithm developed by Shor~\cite{ShorAlgorithm, Shor1997} outperforms the most efficient known classical algorithms.
However, its practical implementation still poses significant challenges on existing quantum computing hardware. A key component of Shor's algorithm is the period-finding subroutine. 
We propose a benchmark protocol based on a related task: finding the period $2^{n}-1$ of a maximum-cycle linear permutation on $n$-long bitstrings.
For special maximum-cycle linear permutations, which are called \emph{maximum-length linear-feedback shift registers}, we find a significant reduction of gate count, as compared to the case of factoring, which makes it more suitable for benchmarking in the present era of quantum computing.
%Furthermore, the evaluation of whether the calculated factors are in fact factors of the investigated number is classically efficient. Finally, the structure of the factoring algorithm employing the quantum Fourier transform (QFT)~\cite{coppersmith2002approximate, nielsen2001quantum} also has similarities to quantum chemistry calculations, highlighting that progress on the implementation of Shor's algorithm can be directly connected to progress for other fields. 
For details of the protocol, see Sec.~\ref{sec:period-finding-benchmark}.

\emph{Quantum error correction benefit:} Any hardware is prone to errors. Quantum computer limitations in this regard can be overcome by quantum error correction~\cite{Terhal2015, Roffe03072019}, where quantum information is redundantly encoded across several physical qubits. While there is an overhead in resources, the encoded information should result in reduced error rates. The benchmark, described in detail in Sec.~\ref{sec:qec-benchmark}, quantifies the benefit of using encoded over unprotected quantum information.

%Taking these considerations into account,  we have developed benchmarks for the Quantum Flagship program of the European Union as a suite of key performance indicators (KPIs), which comply with the following set of requirements:
%\begin{enumerate}
%    \item They are hardware agnostic.
%    \vspace{-1ex}
%    \item They address aspects of the performance of a quantum computer \textit{system}, not a single component.
%    \vspace{-1ex}
%    \item They are \textit{volumetric} in nature, in the sense that they grow with the size of the quantum computer as long as other quantifiers such as connectivity and fidelity grow correspondingly.
%    \vspace{-1ex}
%    \item They are not meant to be driven by use cases. Albeit this would be the ultimate goal, they rather benchmark routines and elements that are typically used in relevant use cases. Actual use cases often require strong downselection to hard instances, which makes the benchmarking procedure more cumbersome.
%    \vspace{-1ex}
%    \item Even though we expect that these benchmarks primarily will be used for quantum-to-quantum comparisons, we focus on cases where quantum computers are expected to outperform classical technologies.
%\end{enumerate}

\section{Clifford Volume benchmark}
\label{sec:clifford-volume-benchmark}

The Clifford Volume benchmark is based on the implementation of randomly chosen $N$-qubit Clifford unitaries as a suitable problem set. It aims to assess the overall computing capacity of the processor and can be understood as a scalable alternative for determining the processor's ``quantum computing volume"---the largest task it can solve within acceptable error limits.

\subsection{Background}

The Clifford group consists of the set of unitaries that map tensor products of Pauli matrices to other tensor products of Pauli matrices through conjugation. According to the Gottesman--Knill theorem, a quantum circuit composed entirely of stabilizer operations---including Clifford gates, the preparation of stabilizer states, and measurements in the computational basis 
(or in an equivalent basis related to the stabilizers)---can be efficiently simulated on a classical computer (see recent numerical methods in Refs.~\cite{1,Gidney_STIM}). Furthermore, an $N$-qubit stabilizer state can be validated using single-qubit Clifford gates and measurements in the computational basis. This validation strategy involves estimating the expectation values of Pauli-string observables corresponding to the stabilizer operators of the state. Based on this validation scheme, an explicit expression for the minimum fidelity between the stabilizer state and any other state generated by the (potentially faulty) implementation of the Clifford unitary can be derived. This worst-case fidelity can be certified with high probability using only $O(N^3)$ measurements (see related works on validating stabilizer states~\cite{PhysRevA.99.042337}). Unlike quantum state tomography, which scales exponentially with the number of qubits, this method remains feasible and efficient even for large systems, ensuring its practicality as a benchmarking tool. 

There exist methods to assess the performance of quantum processors based on the application of layers of single- and two-qubit Clifford gates~\cite{Chen_2023,Merkel2025CliffordProxy}. Here we opt for a more versatile approach by focusing on randomly selected $N$-qubit Clifford operators, which can then be compiled specifically for the platform on which they are implemented. We deliberately avoid prescribing any specific compilation rules that could favor systems with particular connectivity or predefined characteristics. Instead, the chosen Clifford unitaries can be decomposed and optimized during transpilation to the native gates and specifications to the given platforms, thereby enabling the highest possible performance during the benchmarking process~\cite{CliffordVolume2025}. 

%The Clifford group consists of the set of unitaries that map tensor products of Pauli matrices to other tensor products of Pauli matrices through conjugation. According to the Gottesman–Knill theorem, starting from any computational basis state and applying quantum circuits representing a Clifford unitary, the computational basis measurements can be efficiently simulated on a classical computer (see recent numerical methods in Refs.~\cite{1,5}), hence the scalable property of the benchmark. 

\subsection{The benchmark}

In order for the benchmarking task to be manageable in practice, while at the same time ensuring that a given platform cannot be optimized to the implementation of single specific Clifford unitaries, we propose that a minimum number of four uniformly sampled random elements from the $N$-qubit Clifford group be implemented and executed for a given value of $N$. (For an efficient sampling method of Clifford operations, see Ref.~\cite{4}.) Assessing the quality of the implementation of these few, but randomly chosen unitaries, already provides sufficient insight into the achievable overall performance of the hardware.  

For the evaluation of the quality of the individual $N$-qubit Clifford unitary realizations, we propose to use a metric based on the measured expectation values of a certain number of randomly chosen $N$-qubit Pauli operators (stabilizers as well as non-stabilizers, corresponding to the given Clifford unitary). By analyzing the expectation values of these operators and their standard deviations, we can effectively assess the ``correctness" of the implemented Clifford operator. One key advantage of this approach is that it avoids the need for reconstructing experimental output probabilities, a process that requires an exponential number of measurements in the number of qubits. Additionally, the measurement of expectation values is a fundamental capability of quantum processors, aligning with the demands of many real-world applications.

\subsection{Step-by-step description of the benchmark protocol}

For an efficient practical implementation of the Clifford Volume benchmark it is advantageous to  determine the candidate values of $N$ for which the CLV protocol shall be carried out. A possible strategy for this purpose is to apply an adaptive search over the
qubit number $N$, for example using binary search, or a suitable strategy based on individual trials. 
One may also apply numerical simulations using precise, device-specific error models to locate the interval where the crossing of the threshold values can be expected. For a given value of $N$ in the identified interval, the benchmark shall be evaluated by executing the steps outlined below:
\begin{enumerate}
\item \textbf{Initialization:} 
\begin{enumerate}
\item Set the value of $N$. % defining the value of the Clifford group size. 
\item Generate four randomly sampled $N$-qubit Clifford unitaries $\mathcal{G}^m$, $m \in \left[1,4 \right]$, selected without replacement. For each $\mathcal{G}^m$, choose four generators $\mathcal{S}^m_i$, $i \in \left[1,4 \right]$, randomly from the stabilizer group corresponding to the state prepared by the random Clifford operation. Additionally, choose four distinct $N$-qubit Pauli operators $\mathcal{D}^m_i$, $i \in \left[1,4 \right]$, randomly, which lie outside the stabilizer group.
\end{enumerate}

\item \textbf{Circuit preparation:} \\
Construct a circuit implementation for each $\mathcal{G}^m$. Compile each circuit according to the architecture of the quantum computer being benchmarked.

\textbf{Note:} %(\textit{Depth of Clifford operators in LNN topologies}): 
In the linear nearest-neighbor architecture, the implementation of an arbitrary Clifford circuit is possible with a two-qubit depth of only $9N$~\cite{4}.  

\item \textbf{Circuit execution:} 

\textit{This step must be carried out on a quantum device.}

For each $\mathcal{G}^m$, measure the expectation value of each corresponding Pauli operator $\mathcal{S}^m_i$ and $\mathcal{D}^m_i$ by implementing the following steps:

\item[] For $m$ from $1$ to $4$:
    \begin{enumerate}
        \item[] For $i$ from $1$ to $4$:
        \begin{enumerate}
            \item[] For $l$ from $1$ to $L$ (can be interpreted as the number of shots, with a minimum value of 512):
             \begin{enumerate}
                 \item[(a)] Initialize the system in the $\left|0\right\rangle^{\otimes N}$ state.
                 \item[(b)] Execute the quantum circuit corresponding to $\mathcal{G}_m$ and measure the  stabilizer generator $\mathcal{S}^m_i$.
             \end{enumerate}
             \vspace{0.5ex}
            \item[] For $l$ from $1$ to $L$:
             \begin{enumerate}
                 \item[(a)] Initialize the system in the $\left|0\right\rangle^{\otimes N}$ state.
                 \item[(b)] Execute the quantum circuit corresponding to $\mathcal{G}_m$ and measure the  Pauli operator $\mathcal{D}^m_i$.
             \end{enumerate}
             \vspace{0.5ex}
             \item[] Calculate the expectation values $ \left\langle \mathcal{S}^m_i \right\rangle$ and $\left\langle \mathcal{D}^m_i \right\rangle$ from the measurement results and the corresponding standard deviations. % of the stabilizer generator $\mathcal{S}^m_i$, and the Pauli operator $\mathcal{P}^m_i$, respectively.
            \end{enumerate}
        \end{enumerate}

    \textbf{Notes:}
    \begin{enumerate}
        \item For a quantum device capable of performing measurements in the standard computational basis only, the required measurements above can be achieved by completing the circuit implementing the Clifford operation $\mathcal{G}_m$ with a layer of single-qubit Clifford operations before the measurements in the computational basis.
        \item When performing the benchmark, the qubits are always initialized in the $\ket{0}^{\otimes N}$ state. One might think that this approach could lead to certain errors being overlooked—specifically, those that manifest only when the qubits are in the $\ket{0}^{\otimes N}$ state. To address this potential issue, one could initialize the system in a different, randomly selected stabilizer state. However, since we already use uniformly sampled Clifford unitaries for the benchmark, which inherently create a random stabilizer state, it is unnecessary to prepare the system in a different stabilizer state beforehand. Moreover, doing so would significantly increase the complexity of the benchmark circuit.
        \item To reduce measurement bias, we propose applying a layer of single-qubit $X$ gates before the measurements.
    \end{enumerate}

\item \textbf{Performance evaluation:} \\
If the benchmarked quantum processor possesses sufficiently high computational capacity, it should reliably differentiate between stabilizer and non-stabilizer Pauli operators based on the measured expectation values.
    
    To validate this differentiation, one can analyze the distribution of the measured expectation values. Ideally, in an error-free scenario, the expectation value would be $1$ for any Pauli operator within the stabilizer group and $0$ for any non-stabilizer Pauli operator (sometimes called destabilizers). Consequently, it is crucial that the measured expectation values, as well as their averages, meet the following criteria: the expectation values corresponding to stabilizer operators should be significantly greater than $0$, while those associated with non-stabilizer operators should remain approximately $0$, with an appropriate level of accuracy.

    These conditions can be formulated in the following way. We consider the benchmark successful if all the individual stabilizer expectation values are above $1/e$ with $2\sigma_{\mathcal{S}}$ confidence, i.e.,
    $$
    \left\langle \mathcal{S}^m_i \right\rangle
    -
    2 \sigma^{m}_{S_{i}}
    \geq
    \frac{1}{e}
    \approx
    0.3679,
    \qquad
    \forall\, i,m .
    $$
    Here
    $$
    \sigma^{m}_{\mathcal{S}_{i}}
    =
    \sqrt{\frac{1 - \left\langle \mathcal{S}^m_i \right\rangle^2}{L}}
    $$
    is the standard deviation of the measured expectation value due to statistical uncertainty, and $e$ is Euler’s number.
    
    Analogously, for the destabilizer operators, we require that
    $$
    \left|
    \left\langle \mathcal{D}^m_i \right\rangle
    \pm
    2\sigma^{m}_{\mathcal{D}_i}
    \right|
    \leq
    \frac{1}{2e}
    \approx
    0.1839,
    \qquad
    \forall\, i,m ,
    $$
    where
    $$
    \sigma^{m}_{\mathcal{D}_i}
    =
    \sqrt{\frac{1 - \left\langle \mathcal{D}^m_i \right\rangle^2}{L}} .
    $$
    
    Moreover, we also prescribe a criterion on the average performance. Specifically, for every sampled Clifford unitary, we require that the average stabilizer expectation value lies at least five standard deviations above the stabilizer threshold, and that the average destabilizer expectation value lies at least five standard deviations below the destabilizer threshold. Formally, these conditions read
    $$
    \begin{cases}
    \displaystyle
    \overline{\left\langle \mathcal{S}^{\,m} \right\rangle}
    -
    5\,\overline{\sigma}_{\mathcal{S}^{\,m}}
    \ge
    \frac{1}{e},
    \\[3mm]
    \displaystyle
    \left|
    \overline{\left\langle \mathcal{D}^{\,m} \right\rangle}
    \right|
    +
    5\,\overline{\sigma}_{\mathcal{D}^{\,m}}
    \le
    \frac{1}{2e},
    \end{cases}
    \qquad
    \forall\, m .
    $$
    Here
    $$
    \overline{\left\langle \mathcal{S}^{\,m} \right\rangle}
    =
    \frac{1}{4}\sum_{i=1}^{4}
    \left\langle \mathcal{S}^m_i \right\rangle,
    \qquad
    \overline{\left\langle \mathcal{D}^{\,m} \right\rangle}
    =
    \frac{1}{4}\sum_{i=1}^{4}
    \left\langle \mathcal{D}^m_i \right\rangle ,
    $$
    and
    $$
    \overline{\sigma}_{\mathcal{S}^{\,m}}
    =
    \sqrt{\frac{1}{4}\sum_{i=1}^{4}
    \left(\sigma^{m}_{\mathcal{S}_i}\right)^2},
    \qquad
    \overline{\sigma}_{\mathcal{D}^{\,m}}
    =
    \sqrt{\frac{1}{4}\sum_{i=1}^{4}
    \left(\sigma^{m}_{\mathcal{D}_i}\right)^2}.
    $$
    
    \textit{Note}: The choice of $1/e$ for the threshold is motivated by the fact that, under a baseline error model, e.g., a global depolarizing channel with a fixed error probability per circuit layer, the expectation values decay roughly exponentially with circuit depth, and thus, in the optimal case, with the number of qubits. Consequently, reaching $1/e$ ($\sim37\%)$ of the initial expectation value is a natural way to define the effective ``lifetime" of the system.

\end{enumerate}

\subsection{Definition of the benchmark score and reporting}

The largest $N$ value for which the benchmark protocol succeeds serves as a metric to characterize the quantum computer. This benchmark score reflects the number of qubits of sufficient quality required to reliably perform a random Clifford unitary operation on all chosen qubits.

%We require that along with the reported benchmark score, experimental groups/providers also report the ensemble of randomly sampled N-qubit Clifford operations that they implemented when running the benchmark. They should also provide their circuit decompositions of the randomly selected Clifford operations, as well as the randomly chosen non-stabilizer Pauli strings that they measured. Furthermore, they should also report which set of generators of the stabilizer group they chose to measure. 

\subsection{Numerical performance analysis}

\begin{figure}[h]
    \centering
    \includegraphics[width=0.75\linewidth]{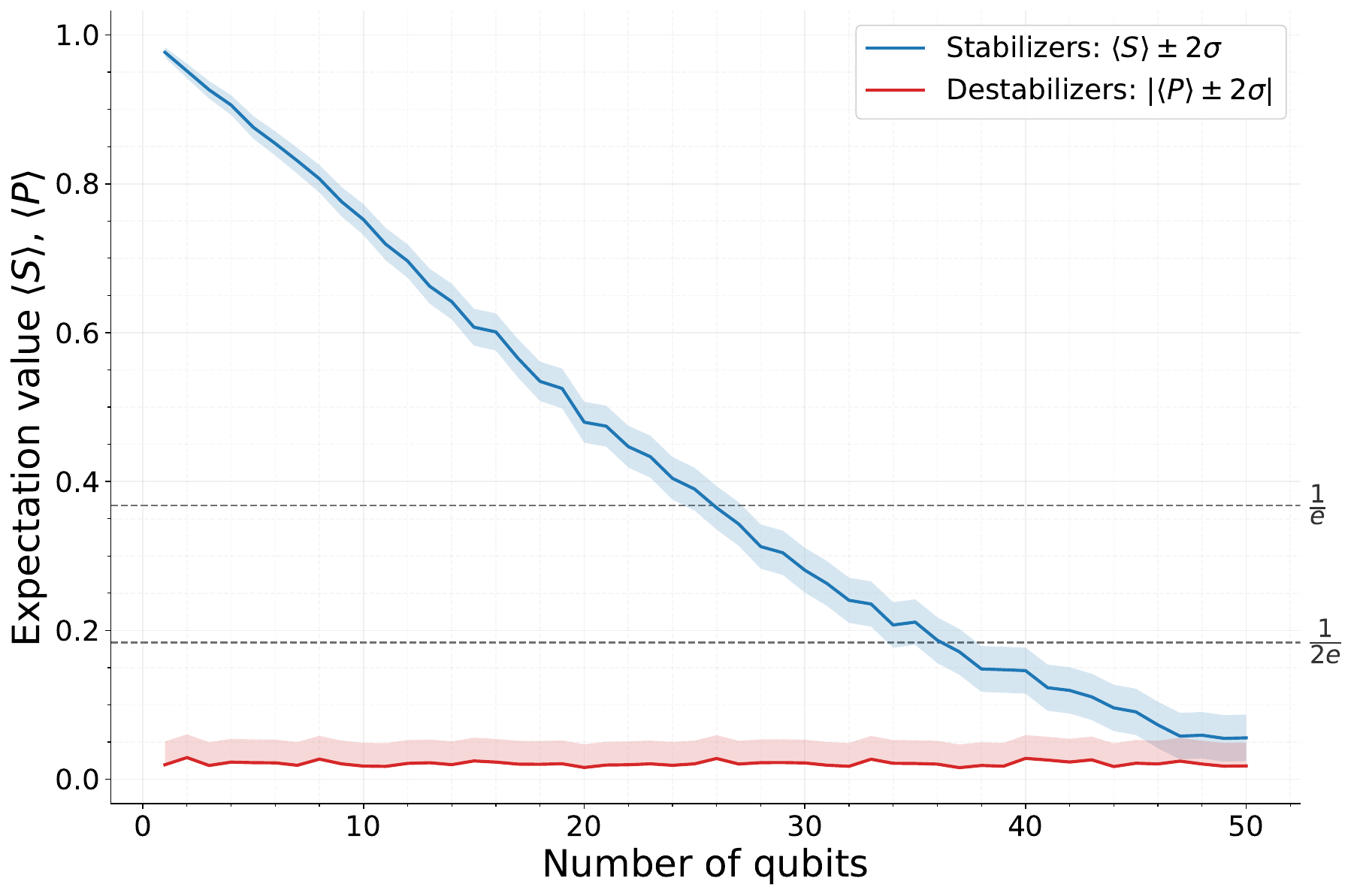}
    \caption{Worst-case expectation values of randomly chosen stabilizer operators (blue curve) and non-stabilizer operators (red curve) corresponding to four randomly chosen Clifford unitaries as a function of the qubit number $N$. For the simulation, error probabilities of $p_{2Q}=10^{-3}$ and $p_{m}=10^{-2}$ were assumed for two-qubit gates and for the readout, respectively. Shaded regions represent the statistical uncertainties of $\pm 2\sigma$.}
    \label{fig:exp_values}
\end{figure}

We have carried out numerical simulations to show how the expectation values of randomly chosen stabilizer and non-stabilizer operators corresponding to $N$-qubit Clifford operators behave in the presence of two-qubit gate errors and readout errors as the number of qubits $N$ grows. The results from these simulations are shown in Fig.~\ref{fig:exp_values}. It can be seen there that as the number of qubits (and consequently, the size of the Clifford unitary) grows, the expectation values of stabilizer operators gradually decrease, due to the errors, until they eventually cross the $1/e$ threshold. The expectation values of the non-stabilizer operators, however, only fluctuate around the value 0.

\begin{figure}[h]
    \centering
    \includegraphics[width=0.55\linewidth]{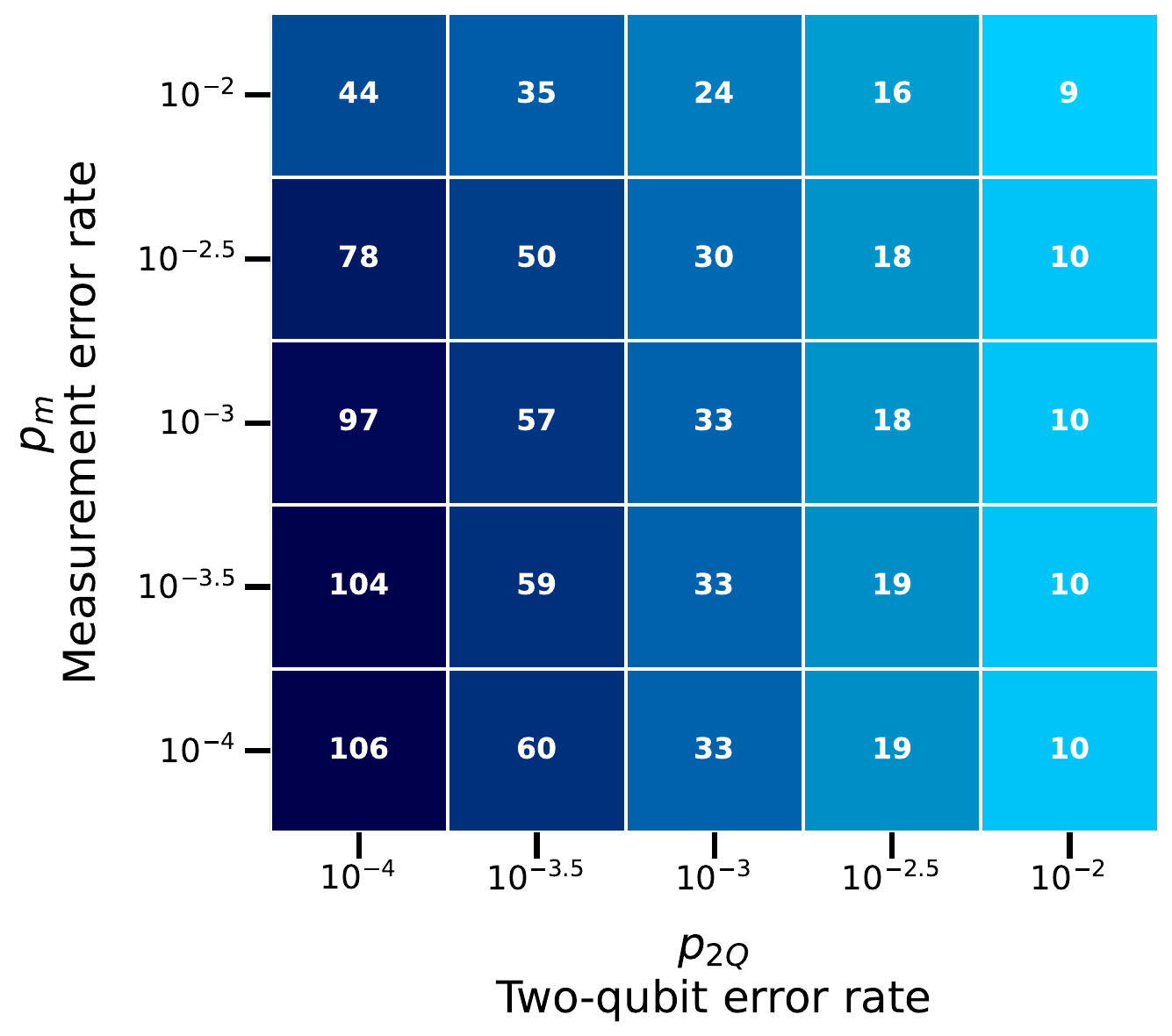}
    \caption{Clifford Volume benchmark scores for different pairs of error parameters. The horizontal axis corresponds to the two-qubit gate error probability $p_{2Q}$, and the vertical axis to the readout error probability $p_m$. Each cell indicates the largest qubit count $N$ for which the benchmark criteria are satisfied under the corresponding noise configuration.}
    \label{fig:CLV_values}
\end{figure}

The results of a more extensive analysis for the CLV scores achievable for a given $\left(p_{2Q},p_{m}\right)$ pair of two-qubit gate error and readout error probability are shown in Fig.~\ref{fig:CLV_values}. For the numerical simulations, we utilized an implementation of the Clifford Volume benchmark protocol together with the corresponding evaluation tools provided in the open-source repository ~\cite{EQCB}.
 
\subsection{Remarks}

\begin{enumerate}

%\item \textit{Reporting on the details of the experimental implementations:} We require that along with the reported benchmark score, experimental groups/providers also report the ensemble of randomly sampled N-qubit Clifford operations that they implemented when running the benchmark. They should also provide their circuit decompositions of the randomly selected Clifford operations, as well as the randomly chosen non-stabilizer Pauli strings that they measured. Furthermore, they should also report which set of generators of the stabilizer group they chose to measure. 

%\item \textit{Late-NISQ quanutm processors:} We anticipate that more advanced, late-NISQ quantum devices,  will achieve sufficiently low gate and readout error rates. Therefore their performance can be evaluated by using stricter metrics. In this context, one can compute the stabilizer fidelity witness (introduced in Ref.~\cite{PhysRevA.99.042337}) as follows:
%$$
%F_{\text{min}}^m = 1 - \sum_{i=1}^N \frac{1 - \langle \mathcal{S}_i^m \rangle}{2},
%$$
%for each $m$. In this scenario, the operation of the quantum computer would be considered successful if the average fidelity witness exceeds $0.5$:

%$$
%F_{\text{min}}^{\text{avg}} = \frac{1}{M} \sum_{m=1}^M F_{\text{min}}^m = 1 - \frac{1}{M} \sum_{m=1}^M \sum_{i=1}^N \frac{1 - \langle \mathcal{S}_i^m \rangle}{2} > 0.5 \, .
%$$

\item \textit{Integrating multiple approaches for robust quantum benchmarking}: We note that it may be beneficial to use not just a single method to characterize the accuracy of circuit implementation, but to combine multiple approaches for a more comprehensive assessment. For example, benchmarks based on mirror circuits~\cite{Proctor2021,Proctor_2022} address the problem from a different perspective. These benchmarks involve implementing the inverse of the original circuit, which simplifies the verification of the final result. However, due to their symmetric structure, mirror-circuit benchmarks can be insensitive to certain types of errors. The benchmark we presented here is capable of revealing these errors, offering a more thorough evaluation from this point of view. %As a future step, an approach where the Clifford Volume is combined with the mirror-circuit benchmark could also be investigated.

\item  \textit{Readout errors:} We do not allow standard readout error mitigation done through postprocessing of the expectation values. Naturally, increasing the readout fidelity by the use of ancilla qubits is possible.
%In order to enhance the precision of the estimated expectation values, readout error mitigation techniques can be employed. However, in this case, both the results with and without error mitigation must be reported. The results without mitigation will be considered for comparing the capabilities of the device with other quantum processors.
\end{enumerate}

\section{Multipartite entanglement benchmark---GHZ-state preparation}
\label{sec:ghz-benchmark}

%\bigskip
%\normalsize Prepared by Benchmark Teams of OpenSuperQPlus\footnote{Authors: R. Peña, D. Aguirre, A. Acuaviva and M. Sanz \\ [-1mm] 
%Contributors: A. Frisk Kockum,  A. Portik, F. Wilhelm-Mauch, and Z. Zimborás} and  Millenion\footnote{Contributors: A. Erhard, C. Marciniak, T. Monz}  \\

%Author 1$^{1}$, Author 2$^{2,1}$, Author 3$^{2}$, Author 4$^{1}$ \\
\medskip
%\small $^{1}$\textit{Institute 1} \\
%$^{2}$\textit{Institute 2}\\
%\medskip  

\normalsize
Entanglement is a fundamental resource for achieving an advantage in quantum computation. This benchmark evaluates a quantum computer's capability to generate entangled states. Specifically, it measures the largest Greenberger--Horne--Zeilinger (GHZ) state with genuine multipartite entanglement that the provider can reliably generate on the processor. \\

%--------------------------
\subsection{Background}

Entanglement is a resource of quantum mechanics that plays a crucial role in quantum computing~\cite{Horodecki2009}. This resource is essential for a quantum algorithm to achieve a speed-up over classical computation~\cite{Entanglement}. In fact, it has been shown that quantum systems with small entanglement can be efficiently simulated on a classical computer~\cite{PhysRevLett.91.147902,Verstraete2008}. Therefore, the size of the largest genuinely entangled state that can be created is a reasonable metric for characterizing the quality of a quantum computer.
The GHZ state~\cite{Greenberger1989} is a type of maximally entangled quantum state that provides an important resource for numerous applications in quantum metrology~\cite{RevModPhys.90.035005}, quantum teleportation~\cite{QuantumTeleportation}, quantum cryptography~\cite{Man2006}, and error-correcting codes~\cite{Knill2005}. Due to technological advances, GHZ states have been generated on several types of quantum hardware, e.g., trapped ions~\cite{Moses2023,Monz2011}, superconducting circuits~\cite{Song2019,PhysRevLett.119.180511,Barends2014SuperconductingQC,kam2023generationpreservationlargeentangled}, Rydberg-atom arrays~\cite{Omran2019,Graham2022}, photonic~\cite{Zhong2018,PhysRevLett.117.210502}, and silicon-based quantum computers~\cite{Takeda2022}. 
%\begin{figure}[ht]
%    \centering 
    %\includegraphics[width=0.37\columnwidth]{figures/GHZ.pdf}
    %\caption{An example of a quantum circuit for generating $N$-qubit GHZ states, where the circuit depth scales linearly with the number of qubits. %This circuit involves the following steps: First, a Hadamard gate is applied to the first qubit. Subsequently, CNOT gates are applied consecutively, using the first qubit as the control qubit and each of the remaining qubits as target qubits. In this case, the initial state of the quantum computer is the $N$-qubit ground state $|\psi_0\rangle=|0\rangle ^{\otimes N}$.}
%    \label{GHZ}
%\end{figure}

The $N$-qubit GHZ state is
\begin{equation*}
|{\text{GHZ}_N}\rangle=\frac{1}{\sqrt{2}}(|0\rangle^{\otimes  N} + |1\rangle^{\otimes N} ).
\end{equation*}
%where $N$ is the number of qubits of the quantum computer.
There exist various different algorithms to generate GHZ states~\cite{Ozaeta2019,Cruz2019,Blatt2008,Moses2023,Mooney2021}. %Some algorithms are more efficient than others. 
However, the choice of algorithm for efficiently generating the GHZ state largely depends on the characteristics of the quantum computer running the algorithm, i.e., the native gates and the topology of the processor. %In Fig.~\ref{GHZ}, it is shown a straightforward circuit to generate GHZ states where the depth of the circuit scale linearly with the number of qubits. However, this method is inefficient for generating a GHZ state and also requires one qubit with a huge connectivity. In contrast, Fig.~\ref{GHZ_2} shows an alternative efficient method to generate GHZ states, where the depth of the circuit scales logarithmically with the number of qubits. This alternative represents a significantly improved method for generating GHZ states compared to the circuit shown in Fig~\ref{GHZ}. Additionally, in this case, the connectivity required does not scale linearly with the number of qubits. We stress that in both cases the quantum computer is initialized in the $N$-qubit $|\psi_0\rangle=|0\rangle ^{\otimes N}$ state. 
%\begin{figure}[ht]
 %   \centering 
 %   \includegraphics[width=0.92\columnwidth]{figures/GHZ_2.pdf}
  %  \caption{Figure shows an alternative quantum circuit for generating $N$-qubit GHZ states where the circuit depth scales logarithmically with the number of qubits.}
   % \label{GHZ_2}
%\end{figure}

The quantum state fidelity allows us to quantify how close two quantum states are. Therefore, it can be used to evaluate how accurately a quantum state has been generated. Particularly, the fidelity between the GHZ state and the prepared state $\hat{\rho}$ is given by
\begin{equation}
\mathcal{F}=\langle {\text{GHZ}_N} |\hat{\rho}|{\text{GHZ}_N}\rangle= \frac{1}{2}(P + C),
\end{equation}
where 
\begin{equation}
\begin{aligned}
P &= \mathrm{Tr}\!\left(\hat{\rho}\,|0\rangle\langle 0|^{\otimes N}\right)
  + \mathrm{Tr}\!\left(\hat{\rho}\,|1\rangle\langle 1|^{\otimes N}\right),\\
C &= \mathrm{Tr}\!\left(\hat{\rho}\left(|0\rangle\langle 1|^{\otimes N}
  + |1\rangle\langle 0|^{\otimes N}\right)\right).
\end{aligned}
\end{equation}
Here, $P$ is the sum of the probabilities of obtaining the states $|0\rangle ^{\otimes N}$ and $|1\rangle ^{\otimes N}$ (and corresponds to the respective Z-stabiliser), while $C$ is the coherence between the states $|0\rangle ^{\otimes N}$ and $|1\rangle ^{\otimes N}$ (and corresponds to the respective X-stabiliser). 
The fidelity reaches a value equal to $1$ if $\hat{\rho}$ is identical to $|{\text{GHZ}_N}\rangle \langle {\text{GHZ}_N}|$. In the absence of noise, the density matrix should only consist of four terms: two diagonal matrix elements, corresponding to the populations of $|0\rangle ^{\otimes N}$ and $|1\rangle ^{\otimes N}$, and two off-diagonal matrix elements, corresponding to the coherence. The term $P$ can be directly measured in the qubit basis, while $C$ can be measured through experimental methods such as parity oscillations~\cite{Leibfried2005,Monz2011}, multiple quantum coherences (MQC)~\cite{PhysRevA.101.032343,kam2023generationpreservationlargeentangled}, quantum state tomography~\cite{PhysRevLett.119.180511}, and classical shadow tomography \cite{Huang2020}. 

The method for calculating GHZ coherences using MQC is explained in more detail in Ref.~\cite{kam2023generationpreservationlargeentangled}. Although the MQC method has been successfully employed, for some platforms, it is not the most practical approach to estimate the fidelity, especially as the number of qubits increases. Full quantum state tomography requires an exponential number of measurements for a complete description of $N$-qubit states, making it impractical for quantum computers with a large number of qubits. Additionally, an efficient method has been proposed, known as classical shadow tomography, using random Clifford measurements~\cite{Huang2020}. This method constructs an approximate description of a quantum state using a limited number of measurements when detailed knowledge of the entire density matrix is not required. Despite its advantages, this method involves running random Clifford circuits with a depth that scales linearly with the number of qubits, which remains experimentally challenging. 

For a method to be considered practical, it should estimate fidelity in a realistic and viable time, thus facilitating its practical applicability~\cite{David}. Recently, Huang \textit{et al.} proposed a quantity, the so-called shadow overlap~\cite{huang2024certifyingquantumstatessinglequbit}, to certify that a given state $\rho$ is close to $|\text{GHZ}_N\rangle\langle\text{GHZ}_N|$. This quantity can be used to efficiently certify the fidelity $\mathcal{F}$ using a number of measurements that scales quadratically with the number of qubits. The shadow overlap does not directly estimate fidelity, but instead provides a method to certify a lower and upper bound on the fidelity with a target state. An alternative approach that directly estimates the fidelity $\mathcal{F}$ is Direct Fidelity Estimation (DFE) \cite{Flammia_2011}. This approach constructs a fidelity estimator by randomly sampling Pauli operators from the $N$-qubit Pauli group according to a probability distribution determined by the target state. In particular, for the GHZ state, this method requires a number of measurements that depends on the target precision of the fidelity estimate, but not on the number of qubits. Overall, several methods are available to assess the generation of GHZ states on a quantum computer, and the most efficient choice depends on the desired precision $\varepsilon$, the desired confidence level $1-\delta$, and the number of qubits involved in the state preparation.

The multipartite entanglement benchmark does not specify a method to estimate or certify a lower bound on the fidelity, thereby allowing experimental groups or providers to choose the approach they consider most appropriate. However, it is important to report the method employed. In the following, we briefly discuss the methods of shadow overlap and direct fidelity estimation.

The shadow-overlap protocol is straightforward. For the GHZ state specifically, we begin by applying the Hadamard gate to each qubit: $H^{\otimes  N} |{\text{GHZ}_N}\rangle$. Next, we proceed with the measurements. First, we randomly choose two qubits $\{k_1,k_2\}$, then measure all qubits of the state $\hat{\rho}$ except for qubits $\{k_1,k_2\}$ in the Pauli $Z$ basis. Denote the measurement outcomes collectively by $z$ $\in$ $\{0,1\}^{N-2}$. Finally, qubits $\{k_1,k_2\}$ are measured randomly in the $X$, $Y$, or $Z$ basis.  Denote the post-measurement state of the qubits $\{k_1,k_2\}$ by $\{|s_1\rangle, |s_2\rangle\}$, respectively. Using the collected data, we can calculate the overlap $\omega$, defined as
\begin{equation*}
\omega = {\rm Tr}(L_{zk} \sigma),
\end{equation*}
where
%\begin{equation*}
\begin{align}
L_{zk}&= \sum_{ \substack{l_1,l_2 \in \{0,1\}^{r} \\ {\rm dist}(l_1,l_2)=r}} |\Psi_{z_k}^{l_1,l_2}\rangle \langle \Psi_{z_k}^{l_1,l_2}| ,\\
|\Psi_{z_k}^{l_1,l_2}\rangle &=\frac{\Psi(z_k^{(l_1)})|l_1\rangle + \Psi(z_k^{(l_2)})|l_2\rangle}{\sqrt{|\Psi(z_k^{(l_1)})|^2+|\Psi(z_k^{(l_2)})|^2}},
\\
\sigma &= (3|s_1\rangle \langle s_1| - \mathbb{1}) \otimes (3|s_2\rangle \langle s_2| - \mathbb{1}).
\end{align}
%\end{equation*}
This process is repeated $T$ times to obtain the overlaps $\omega_1,...,\omega_T$. Finally, the shadow overlap $\bar{\omega}$ can be estimated as 
\begin{equation*}
\bar{\omega} = \frac{1}{T} \sum_{t=1}^T \omega_t.
\end{equation*}
For more details, see Ref.~\cite{huang2024certifyingquantumstatessinglequbit}.

On the other hand, DFE provides a method to directly estimate the fidelity. In this approach, the fidelity between the target state $|\psi\rangle$ and the prepared state $\hat{\rho}$ can be written as
\begin{equation}
\mathcal{F}
= \sum_{P \in \mathcal{P}_N} c_P \, \langle P \rangle ,
\end{equation}
where $\mathcal{P}_N$ denotes the $N$-qubit Pauli group,
$\langle P \rangle = \mathrm{Tr}(\hat{\rho}\, P / \sqrt{2^N})$ 
denotes the normalized expectation value of the Pauli operator $P$, and the
coefficients $c_P = 2^{-N/2}\langle \psi | P | \psi \rangle$ are determined by the target state. For the GHZ state, the relevant Pauli operators form a restricted subset of the $N$-qubit Pauli group, which is the subset sampled in DFE~\cite{Flammia_2011}.

%---------------------------
\subsection{The benchmark}
The metric aims to report the largest GHZ state with genuine multipartite entanglement that can be generated on the quantum processor. It provides standardized information about the size of the qubit cluster that can be entangled in a quantum state relevant for various applications, and which is, in principle, both straightforward to generate and measure.

%---------------------------
\subsection{Step-by-step description of the benchmark protocol}
%\begin{figure}[ht]
 %   \centering 
    %\includegraphics[width=1.01\columnwidth]{figures/Diagram.pdf}
    %\caption{Flow chart of the protocol for computing the GHZ metric. %Users must execute the protocol under both base and peak performance conditions and report the results for each scenario.
    %}
    %\label{protocol}
%\end{figure}

For an efficient practical implementation of the benchmark it is advantageous to determine the candidate values of $N$ for which the GHZ state preparation benchmark protocol shall be carried out. Strategies, similar to the ones mentioned in the case of the CLV benchmark can be applied here too: e.g., using binary search, a suitable strategy based on individual trials, or using numerical simulations incorporating device-specific error models. For a given value of $N$ the benchmark shall be evaluated by executing the steps outlined below:

%The procedure shall start by implementing an $N$-qubit GHZ state, i.e., setting $N = N_{\mathrm{total}}/2$, where $N_{\mathrm{total}}$ is the total number of qubits of the quantum processor. The value of $N$ can then be adjusted adaptively depending on whether the prescribed success criterion is satisfied. The threshold can be efficiently located using a binary search strategy, reducing the number of required benchmark executions to logarithmic scaling in $N_{\mathrm{total}}$. For each tested value of $N$, the benchmark can be evaluated by executing the steps outlined below:

\begin{enumerate}
\item \textbf{Circuit construction:} \\
%\begin{itemize}
%\item In the base performance case, start using the circuit shown in Fig.~\ref{GHZ} to generate the GHZ state using $3$ qubits. Although in principle the possibilities of developing the GHZ state using this quantum circuit scale rapidly with the number of qubits, we expect that only the best options will be analyzed, such as qubits that are directly connected to each other and the highest-quality connections. Users must use a common standard transpiler.
The user can design the best circuit they can to generate the GHZ state using $N$ qubits adapted to their processor. Users are allowed to employ any tools and techniques to achieve the most efficient generation of the GHZ state. Consequently, any transpiler may be used.
%\end{itemize}

\item  \textbf{Circuit execution:} \\
Execute the quantum circuit to generate the GHZ state on the quantum computer. Different strategies can be employed to estimate or certify the fidelity between the prepared state and the target GHZ state, each requiring a different number of measurements.

\begin{enumerate}[label=(\alph*)]
\item If one aims to measure $P$ and $C$ to estimate the fidelity, then the number of measurements required for the true fidelity to be $F_{\rm true}\geq F-\varepsilon$, with probability $\mathrm{Pr}\left[F_{\rm true}\geq F-\varepsilon\right]\geq 1-\delta$, is~\cite{PhysRevA.99.042337} 
$$T\approx \mathcal{O}\left( \frac{N^{2}\, \mathrm{log}\left(\frac{1}{\delta}\right)}{\varepsilon^{2}}\right).$$

\item If one employs the shadow-overlap protocol to certify multipartite entanglement,
the number of measurements required is
\begin{equation}
T = 256\,N^2 \log\!\left(\frac{2}{\delta}\right),
\end{equation}
which guarantees that the prepared state
exhibits genuine multipartite entanglement whenever the estimated shadow overlap
exceeds the corresponding threshold~\cite{huang2024certifyingquantumstatessinglequbit}.

\item Finally, DFE provides a direct estimator of the fidelity. The number of measurements required for this is \begin{equation}
T = \left\lceil \frac{8\log(4/\delta)}{\varepsilon^2} \right\rceil.
\end{equation}
\end{enumerate}

%\item \textbf{Compute shadow overlap or fidelity $\mathcal{F}$:} \\ 
%Compute the shadow overlap, or directly estimate the fidelity $\mathcal{F}$.

\item \textbf{Evaluate shadow overlap or fidelity:}

\begin{enumerate}[label=(\alph*)]
\item \textbf{Shadow overlap}:
%\begin{itemize}
 If $\bar{\omega}\geq 1- \frac{3}{4N}$ with $\epsilon < 1/N$ holds, it implies that 
 $$\langle \text{GHZ}_N| \hat{\rho} | \text{GHZ}_N\rangle \geq 1/2.$$
 If the above condition for $\bar{\omega}$ is satisfied with $3\sigma$ confidence, the current value of $N$ is considered successful and the search proceeds by testing a larger value of $N$ according to the chosen search strategy.

\vspace{1ex}
\textit{Note:} If the shadow overlap does not exceed the threshold, one may repeat steps 1 to 3 with a different set of $N$ qubits or estimate the fidelity using an alternative method. %If all sets of $N$ qubits have been tested and the threshold has not been achieved, the GHZ metric will be equal to the largest value of $N$ for which the circuit was successfully executed, i.e., the last successful value prior to failure.

%\end{itemize}

\vspace{1ex}
%\begin{itemize}
\item \textbf{Fidelity}:
%\begin{itemize}
If %the fidelity $\mathcal{F}$ exceeds $1/2$ (
$\mathcal{F} > 1/2$ with $3\sigma$ confidence, indicating genuine multipartite entanglement~\cite{Guhne2010,Leibfried2005}, then repeat steps $1$ to $3$ by testing a larger value of $N$ according to the chosen search strategy.

\vspace{1ex}
\textit{Note:} If the fidelity does not exceed the threshold, one may repeat steps 1 to 3 with a different set of $N$ qubits. %If all sets of $N$ qubits have been tested and the threshold has not been achieved, the GHZ metric will be equal to the largest value of $N$ for which the circuit was successfully executed, i.e., the last successful value prior to failure.
%\end{itemize}
%\end{itemize}
\end{enumerate}
\end{enumerate}

%---------------------------
\subsection{Definition of the benchmark score and reporting}

The GHZ metric is equal to largest value of $N$ for which the GHZ state preparation was successfully executed (i.e., the last successful value prior to failure). %It quantifies the ability of a quantum computer to generate genuine multipartite entanglement is the size of the largest GHZ state with genuine multipartite entanglement, i.e., with fidelity larger than $1/2$~\cite{Guhne2010,Leibfried2005}, that the provider can create. 
This metric provides a lower bound on the fidelity of the prepared GHZ state and, as a consequence, an upper bound on the maximum genuine multipartite entanglement that the quantum computer can successfully generate.

\subsection{Numerical performance analysis}

To illustrate the behavior of the multipartite entanglement benchmark, we carried out numerical simulations using Qiskit Aer~\cite{Javadi-Abhari2024}. The goal of these simulations is to demonstrate how the benchmark score depends on hardware-level noise. We employed a simple noise model. Two-qubit gates were affected by a depolarizing error channel with error probability $p_{2Q}$, applied after each entangling gate. Measurement errors were modeled as independent classical bit-flip errors with probability $p_m$ on each measured qubit. Single-qubit gates were assumed to be ideal, reflecting the fact that their error rates are typically an order of magnitude smaller than those of two-qubit gates on current hardware platforms.

For the GHZ-state preparation, we used a standard logarithmic-depth circuit composed of a single Hadamard gate followed by layers of CNOT gates arranged in a binary-tree structure. Under the assumption of all-to-all qubit connectivity, this construction generates an $N$-qubit GHZ state with circuit depth scaling as $\mathcal{O}(\log N)$, making it well suited for scalable benchmarking and for minimizing unnecessary exposure to noise. On hardware with restricted connectivity, such as square or heavy-hexagon lattices, additional routing is required and the circuit depth instead scales as $\mathcal{O}(\sqrt{N})$.

\begin{figure}
    \centering
    \includegraphics[width=1\linewidth]{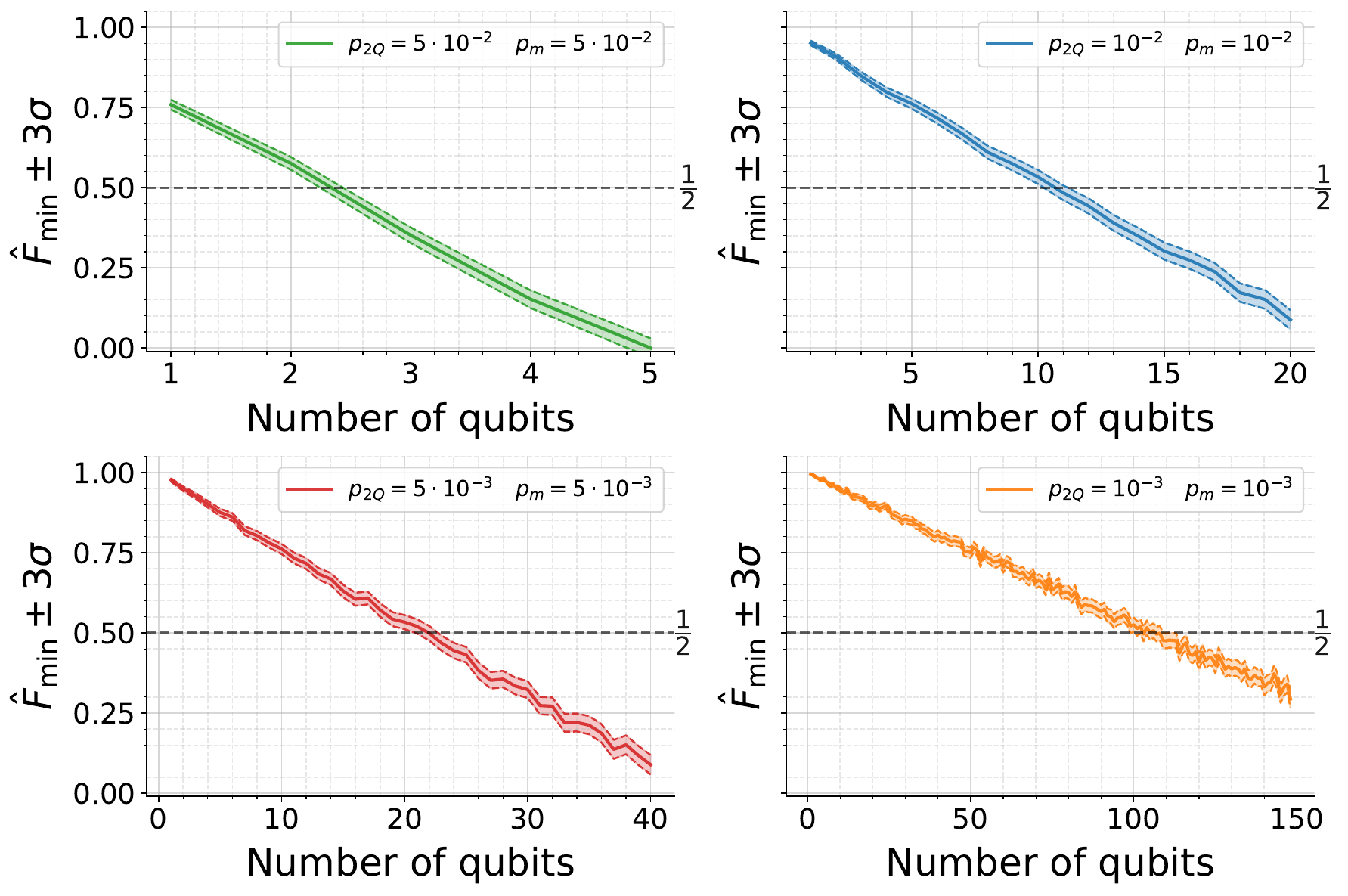}
    \caption{Stabilizer-based lower bound on the GHZ-state fidelity, $\hat F_{\min}$, as a function of the number of qubits for four different combinations of two-qubit depolarizing error probability $p_{2Q}$ and measurement error probability $p_m$. Solid lines show the estimated fidelity value, while shaded regions indicate the $\pm 3\sigma$ statistical uncertainty due to finite sampling. The horizontal dashed line marks the entanglement threshold $\hat F_{\min}=1/2$, above which genuine multipartite entanglement is certified. All expectation values were estimated from $8192$ measurement shots per measurement setting.}
    \label{fig : ghz}
\end{figure}

For each value of $N$, we simulated the full benchmark protocol using a stabilizer-based entanglement witness. The $N$-qubit GHZ state is a stabilizer state and can be uniquely characterized as the simultaneous $+1$ eigenstate of a set of commuting Pauli operators,
$$
\mathcal{G} = \{ X^{\otimes N}, Z_1 Z_2, Z_2 Z_3, \dots, Z_{N-1} Z_N \}.
$$
The expectation values of all stabilizer generators were estimated using only two global measurement settings: measurements of all qubits in the $X$ basis to estimate $\langle X^{\otimes N} \rangle$, and measurements of all qubits in the $Z$ basis to estimate the nearest-neighbor correlators $\langle Z_k Z_{k+1} \rangle$. Importantly, this measurement strategy is independent of the system size $N$ and therefore scalable. From the measured stabilizer expectation values $\tilde{\mu}_l$, we computed a worst-case lower bound on the GHZ-state fidelity consistent with the measurement data,
$$
F_{\min} = \max\!\left( 0,\; 1 - \frac{1}{2} \sum_{l=0}^{N-1} (1 - \tilde{\mu}_l) \right).
$$
This quantity provides an estimate of the true fidelity of the prepared state. We evaluated whether the fidelity exceeded the threshold $F_{\min} > 1/2$, which certifies genuine multipartite entanglement. Statistical uncertainties were estimated from finite sampling, in accordance with the measurement requirements specified in the benchmark protocol.

\begin{figure}
    \centering
    \includegraphics[width=0.55\linewidth]{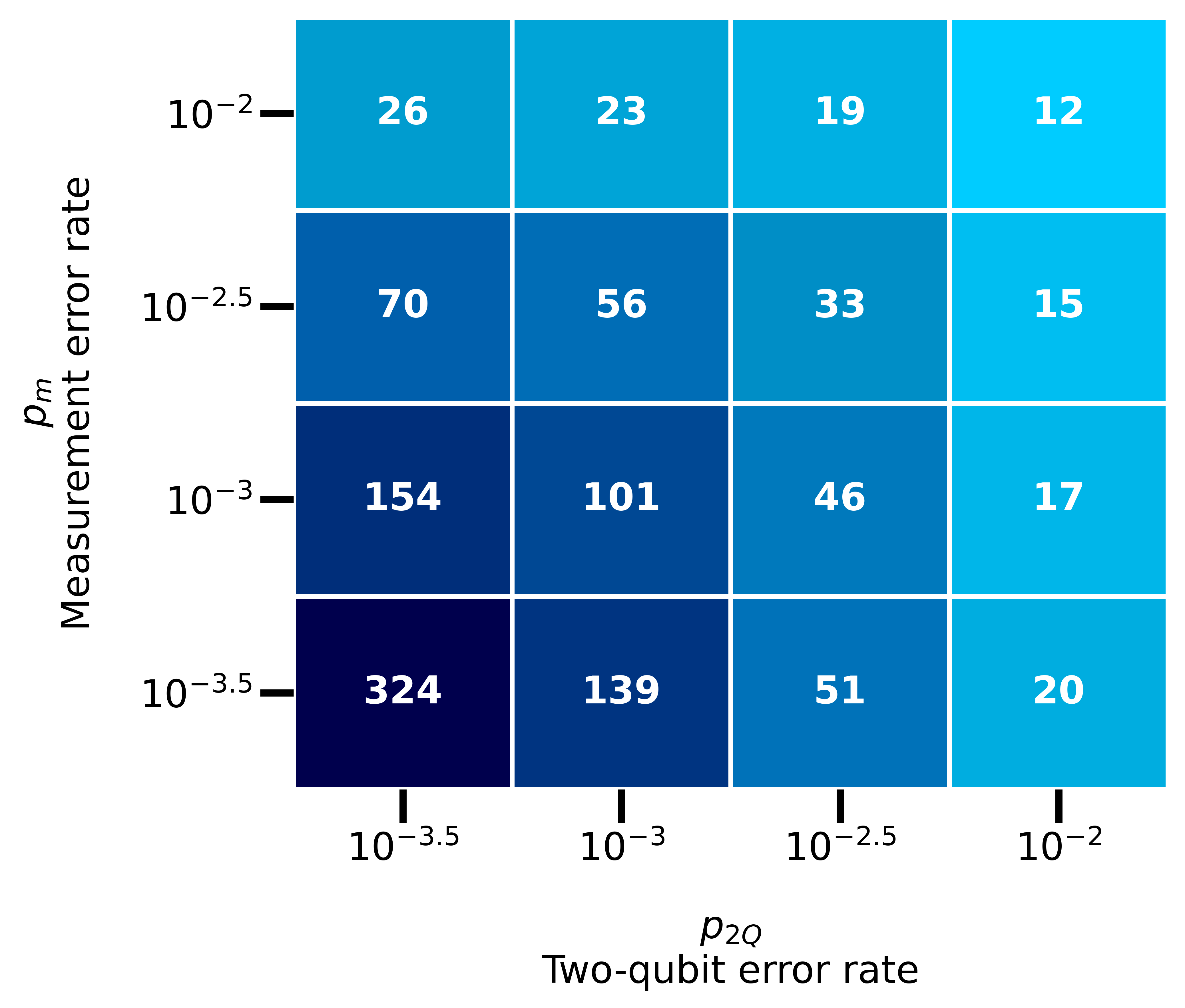}
    \caption{Multipartite entanglement benchmark scores obtained from numerical simulations for different combinations of two-qubit depolarizing error probability $p_{2Q}$ and measurement bit-flip error probability $p_m$. Each cell shows the largest number of qubits $N$ for which the stabilizer-based fidelity lower bound $\hat F_{\min}$ remains above the entanglement threshold $1/2$ with $3\sigma$ statistical confidence. The results illustrate the strong dependence of the certifiable GHZ state size on both gate and measurement error rates.}
    \label{fig:ghzvolume}
\end{figure}

The simulations were performed for a range of qubit numbers and for four different combinations of error parameters $(p_{2Q}, p_m)$, spanning regimes representative of current and near-term quantum hardware. As expected, the GHZ fidelity decreases with increasing system size due to the accumulation of gate and measurement errors (see Fig.~\ref{fig : ghz}). For sufficiently small error rates, the stabilizer-based fidelity bound remains above the entanglement threshold for larger values of $N$, whereas higher error rates lead to an earlier breakdown of certified multipartite entanglement. These results illustrate how the GHZ benchmark provides a clear and scalable indicator of a quantum processor’s ability to generate and maintain genuine multipartite entanglement under realistic noise conditions (see Fig.~\ref{fig:ghzvolume}, which shows the benchmark scores obtained in simulations for a larger spread of error parameters). The numerical simulations were carried out using an open-source implementation of the multipartite GHZ entanglement benchmark, with the corresponding evaluation tools provided in ~\cite{EQCB}.

%---------------------------
\subsection{Remarks}
There are multiple remarks and clarifications about how this metric should be understood and about its limitations.

\bnm
\item \textit{Fair comparison}: To enable a fair comparison between different quantum computers, it is crucial that each technology competes based on its own merits, without introducing biases or artificial advantages/limitations unrelated to the processor itself, such as the quality or aggressiveness of the transpiler~\cite{David}. Traditionally, to ensure fairness, users report both {\it peak} and {\it base} performance results. Base performance refers to all users adhering to the same transpilation rules for generating the GHZ state. However, in noisy quantum computation, this becomes more complex, as any initial choice of circuit to be transpiled might inherently introduce a bias favoring a specific topology. While this remains an open question, a possible solution could involve defining a family of allowable initial circuits alongside standardized transpilation rules for base performance. For this work, however, we have chosen to report only the peak performance to simplify comparisons. 
\item \textit{Scalability}: From a technical perspective, this metric is not strictly scalable. The reason for this is that if we truly attempt to identify the largest cluster that can be entangled---for example, an \(m\)-qubit cluster on a processor with \(n\) qubits---the number of combinations that must be discarded is strictly \(\binom{n}{m+1}\). If \(m\) is on the order of \(n/2\), this scales exponentially. To avoid this issue, we consider two key points: first, the metric does not refer to the largest cluster that could theoretically be generated but rather the largest cluster that the user/provider can successfully generate. Second, while the combinatorial count mentioned above is strictly correct, it is not meaningful to consider disconnected qubit clusters. Thus, the most reasonable strategy is to incrementally add one qubit to the previously verified cluster. In this way, the number of options scales linearly with the area of the cluster, making the process efficient.

\item \textit{Reporting:} Reporting is crucial for the repeatability of the benchmark. Therefore, when reporting the result of a metric, sufficient information must be provided. In particular, for this benchmark, experimental groups or providers are required to submit a detailed report including the quantum circuit used to generate the GHZ state and the set of qubits employed. The transpiled quantum circuit should be provided. The report must also include information about the quantum processor used, the calibration data corresponding to the time of each circuit execution, the compilation configuration and settings applied to each circuit, and the selected benchmark parameters (such as $\delta$ and $\varepsilon$). Additionally, the report must provide the resulting fidelity or the shadow overlap, the number of measurements \(T\), and the GHZ metric.

\edm

\section{Shor's period-finding benchmark}
\label{sec:period-finding-benchmark}

Shor's integer-factoring quantum algorithm~\cite{ShorAlgorithm, Shor1997, Gidney2021howtofactorbit, gidney2025factor2048bitrsa} outperforms the most efficient known classical factoring procedures.
Its core component is Shor's period-finding subroutine~\cite{nielsen2010quantum}.
Here, we propose a benchmark based on this period-finding component; this benchmark is easier to implement on NISQ devices than factoring itself. 
Furthermore, the structure of the period-finding subroutine employing the quantum Fourier transform (QFT)~\cite{coppersmith2002approximate, nielsen2010quantum} has similarities to quantum chemistry calculations, highlighting that progress on the implementation of period finding is relevant for other applications as well.

\subsection{Background}
Shor's algorithm is a cornerstone of quantum computing. 
It is possible to define algorithmic-level, holistic benchmarks based on Shor's algorithm \cite{davis2021benchmarksquantumcomputersshors, Genting2023, Lubinski2023}. 
However, such benchmarks are not suitable to continuously track the development of current quantum computing hardware, because the factoring of small (say, two- and three-digit) numbers already requires more gates and deeper circuits than what can be meaningfully used today. 

Motivated by this difficulty, we here propose an application-driven volumetric quantum computer benchmark based on the period-finding component of Shor's algorithm.
The benchmark we propose can be evaluated with a CNOT count that scales more favorably than those of the circuits of Shor's factoring algorithm. 
In particular, for a quantum computer with all-to-all CNOT connectivity, we identify an implementation with a CNOT gate count that scales with the problem size $n$ (see below) as $12 n^3/\log_2 (n)$. 
This can be compared to, e.g., the CNOT gate count estimate $217 n^3/\log_2 (n)$ of Ref.~\cite{LiuYangYang} for Shor's factoring algorithm for an $n$-bit integer.
The reduced CNOT gate count, compared to Shor's factoring algorithm, makes our benchmark proposal more suited to track progress of quantum computers in the NISQ era.

\subsection{The benchmark}
The benchmarking task is to use the quantum computer to determine the period of maximum-cycle linear permutations, using Shor's period-finding algorithm.
The linear permutations act on $n$-long bitstrings.
We base this benchmark on special linear permutations that have two cycles: a cycle of length 1 (the zero bitstring is mapped to itself, which is true for any linear permutation), and a cycle of length $r = 2^n-1$, which we refer to as the period of the permutation. 
Therefore, the period of such permutations is exponential in $n$, similar to the period of the modular multiplication used in Shor's algorithm. 
The evaluation of the benchmark consists of the experimental evaluation of the success probability of finding $r$ with the quantum computer, for successively increasing values of $n$. 
Informally, the benchmark score is the highest $n$ where the success probability is above a certain threshold.

\subsection{Step-by-step description of the benchmark protocol}

\begin{enumerate}
\item 
\textbf{Initialization:} \\
Take a quantum computer with qubit number $N_\mathrm{tot}$.
Take $n \leq N_\mathrm{tot}$.
Choose a maximum-cycle linear permutation $P_n$ on $n$-long bitstrings, or, equivalently, on $A_n=\{0,1,\dots,2^{n}-1\}$, which maps 0 to 0 and has a cycle of length $2^n-1$.
%The cycle generated by such a permutation is also called a \textit{maximum length sequence}.

\item 
\textbf{Circuit preparation:} \\
Compile Shor's period-finding circuit for this permutation $P_n$.

\item 
\textbf{Circuit execution:} \\
Execute the circuit $10^4$ times, and perform the post-processing procedure that yields the period candidate for each shot. 
A shot is successful if the period candidate equals the period $r = 2^n-1$. 
Denote the fraction of successful shots with $q_{s,n}$.

\item 
\textbf{Performance evaluation:} \\
It is known that on a perfect quantum computer with sufficiently many qubits, the success probability of Shor's period-finding circuit is sufficiently close to 
\begin{equation}
p_{s,n}= \frac{\phi(r)}{r}, 
\end{equation}
where $\phi$ is Euler's totient function. For a permutation with maximum cycle length, the period is $r = 2^n-1$.

The benchmark score $n_s$ is the largest integer such that the quantum computer achieves a success ratio
\begin{equation}
 \eta \equiv \frac{q_{s,n}}{p_{s,n}} > 0.15,
 \label{eq:threshold}
\end{equation}
for all $n\leq n_s$.

\end{enumerate}

\subsection{Definition of the benchmark score}
Find the largest integer $n_s$ such that for all $n \leq n_s$, the success ratio of determining the period of the permutation $P_n$ exceeds 0.15. This number $n_s$ is the benchmark score.

\subsection{Numerical performance analysis}
Here, we illustrate how the score of Shor's period-finding benchmark depends on the quality of a quantum computer. We do this based on a simple estimate of the success ratio as a function of error rates of elementary operations. The benchmark leaves some freedom in how to implement the period-finding quantum circuit; see, e.g., Remark 3 in Sec.~\ref{sec:shor-remarks} below. However, for the purpose of this illustration, we choose a specific implementation, which we briefly introduce in the first part of this section. Then, we estimate the success ratio and the benchmark score.

\begin{figure}
    \centering
    \includegraphics[width=0.9\linewidth]{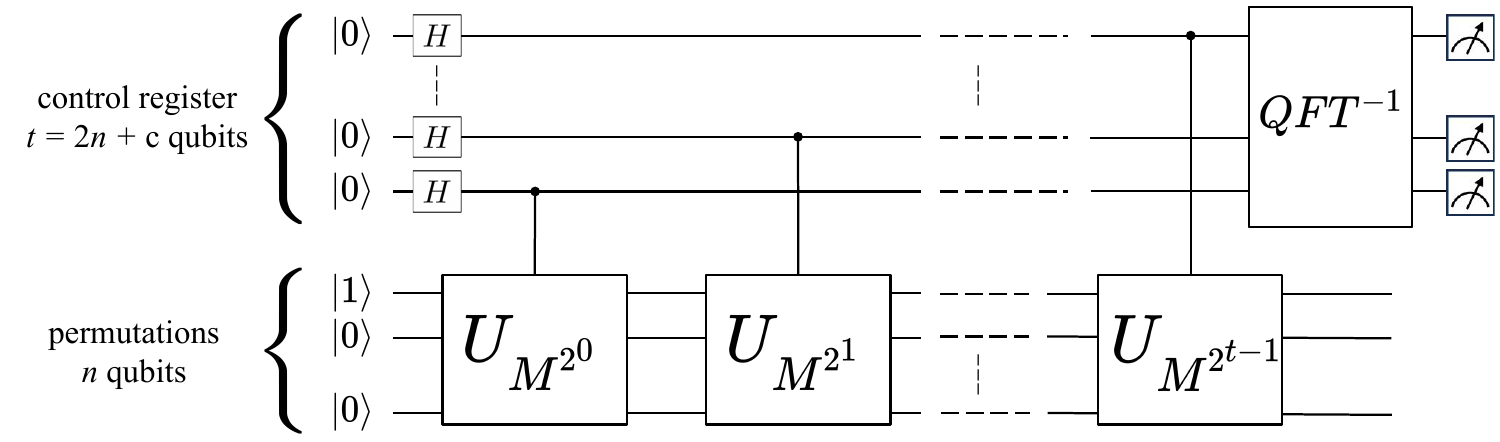}
    \caption{Shor's period-finding quantum circuit. The quantum circuit uses a total of $3n+c$ qubits: $t=2n+c$ qubits in the control register and $n$ qubits to implement  permutation gates. The $U_M$ gate implements a linear permutation specified by the $n\times n$ matrix $M$. Its powers $M^{2^q}$, $q\in \mathbb{N}$ are computed classically and the corresponding quantum gates $U_{M^{2^q}}$ are compiled.}
    \label{fig:shor-circ}
\end{figure}

Shor's period-finding quantum circuit can be implemented as shown in Fig.~\ref{fig:shor-circ}. In the figure, $U_{M}$ denotes the unitary gate implementing a maximum-cycle linear permutation on $n$-long bitstrings, specified by the $n \times n$ binary matrix $M$.
The $i$th entry $b'_i$ of the bitstring $b'$ that results from permuting the bitstring $b$ by $M$ is $b'_i = \left(\sum_{j=1}^n M_{ij} b_j \right) \! \! \mod 2$.

Now, we give a recipe for how such permutations can be found and how to implement the corresponding quantum gates. The first step is to specify a primitive polynomial with binary coefficients $p(x) = x^n + c_{n-1}x^{n-1}+...+ c_0$. Such primitive polynomials are known up to large $n$, even above $4000$~\cite{Zivkovic1994}. The companion matrix of a polynomial of degree $n$ is the $n\times n$ matrix
\begin{equation}
M =
\begin{pmatrix}
    0 & 1 & 0 & \cdots & 0 \\
    0 & 0 & 1 & \cdots & 0 \\
    \vdots & \vdots & \vdots & \ddots & 0 \\
     0 & 0 & 0 & \cdots & 1\\
    c_0 & c_1 & \cdots & c_{n-2} & c_{n-1}
\end{pmatrix}.
\end{equation}
The fact that $p(x)$ is a primitive polynomial guarantees that $M$ specifies a maximum-cycle linear permutation. The quantum circuit in Fig.~\ref{fig:shor-circ} uses controlled gates of the form $cU_{M^{2^q}}$, $q\in \mathbb{N}$. The matrix $M^{2^q}$ can be computed classically using $q$ matrix squaring operations. Note that matrix multiplications are meant modulo 2. Compiling the quantum gate corresponding to the matrix $M^{2^q}$ is possible using $n^2/\log_2 (n)$ CNOT gates~\cite{Patel2008}.

Now we give our estimate for the success ratio and the benchmark score. 
We assume that the elementary operations that are native on the quantum computer are single-qubit gates, CNOT gates, and measurements.
In present-day quantum computers, often the error rates of single-qubit gates are much smaller than those of entangling gates and measurements; therefore, we assume single-qubit gates to be perfect. 
The probability of no error occurring during a run of a quantum circuit is estimated as
\begin{equation}
    P_s(n_{2Q}, n_m) = (1-p_{2Q})^{n_{2Q}} \cdot (1-p_m)^{n_m},
    \label{eq:success-prob}
\end{equation}
where $p_\mathrm{2Q}$ and $p_\mathrm{m}$ are the error rates of CNOT gates and measurements, respectively, while $n_{2Q}$ and $n_m$ are the number of CNOT gates and measurements, respectively, in the circuit. 
We estimate the success probability of finding the period as $p_{s,n}\cdot P_s(n_{2Q}, n_m)$, which yields the following estimate for the success ratio: $\eta \approx P_s(n_{2Q}, n_{m})$.

Next, we determine the scaling of the CNOT gate count and the number of measurements used by the period-finding quantum circuit.
\bnm
\item \emph{The number of qubits to be measured is $2n+1$:}  Shor's period-finding circuit outputs an approximation of $s/r$, where $r$ is the period and $s$ is a random integer uniformly sampled from $\{0, 1,..., r-1\}$. In order to successfully determine $r$, it is sufficient to know $s/r$ accurate to $2n+1$ bits~\cite{nielsen2010quantum}. Therefore, it is sufficient to measure $2n+1$ bits. Note that to ensure that the approximation of $s/r$ is accurate to $2n+1$ bits, the size $t$ of the control register must be slightly larger than $2n+1$, but measuring the first $2n+1$ qubits is sufficient.

\begin{figure}
    \centering
    \includegraphics[width=0.6\linewidth]{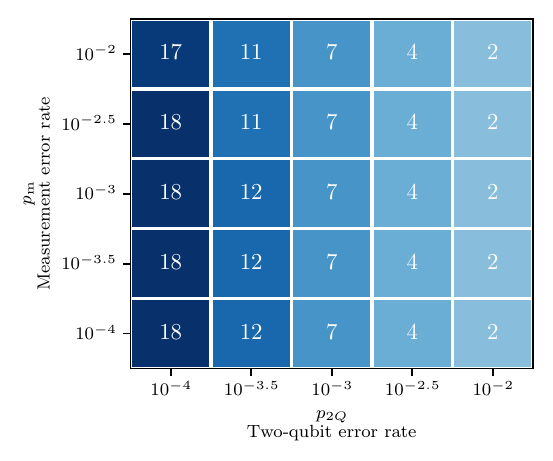}
    \caption{Estimates for the score of Shor's period-finding benchmark. The estimates of the benchmark score based on Eq.~\eqref{eq:success-ratio-est} are displayed for different values of the two-qubit gate error rate $p_\mathrm{2Q}$ and the measurement error rate $p_\mathrm{m}$.
    These estimates correspond to a quantum computer with all-to-all CNOT connectivity.}
    \label{fig:shor-benchmark-score}
\end{figure}

\item \emph{CNOT gate count:} Linear permutations can be compiled using CNOT gates only, and the CNOT gate count in an all-to-all connected device is $n^2/\log_2(n)$ to leading order in $n$~\cite{Patel2008}. For Shor's period-finding algorithm, $2n+c$ controlled linear permutations are needed, which consist of $\sim n^2/\log_2(n)$ Toffoli gates. Each Toffoli can be decomposed into six CNOT gates and single-qubit gates. Therefore, up to leading order, the number of CNOT gates needed is $n_\mathrm{CNOT} = 12n^3/\log_2(n)$. We use this leading-order expression to estimate the number of CNOT gates needed.
\edm
Combining Eq.~\eqref{eq:success-prob} with these gate counts, we obtain the estimate
\begin{equation}
    \tilde{\eta}(n) = (1-p_\mathrm{2Q})^{12n^3/\log_2(n)} \cdot (1-p_\mathrm{m})^{2n+1}.
    \label{eq:success-ratio-est}
\end{equation}
for the success ratio. Note that this is a monotonically decreasing function of $n$. We estimate the benchmark score with the largest $n$ for which $\tilde{\eta}(n)>0.15$. This estimate is shown in Fig.~\ref{fig:shor-benchmark-score}.
Both Eq.~\eqref{eq:success-ratio-est} and Fig.~\ref{fig:shor-benchmark-score} reveal that the benchmark score depends much stronger on the two-qubit error rate than on the measurement error rate.
This is a clear difference compared to the Clifford Volume and GHZ benchmarks.

\subsection{Remarks}
\label{sec:shor-remarks}

\bnm

\item \emph{On the efficiency and practicality of the classical evaluation of this benchmark}:
There are two steps of evaluating the benchmark that scale exponentially with $n$.
\begin{enumerate}
    \item \textit{Finding primitive polynomials to construct maximum-cycle permutations}: Finding primitive polynomials of degree $n$ does not scale efficiently with $n$. However, for all $n < 352$, primitive polynomials are known~\cite{Zivkovic1994}. Therefore, constructing maximum-cycle linear permutations is possible. Primitive polynomials of larger degree are also known but not for all degrees above 352.
    \item \textit{Calculating $\phi(2^n-1)$}: Step 4 of computing the benchmark score requires calculating $\phi(2^n-1)$, which scales exponentially with $n$. However, if the factorization $2^n-1 = p_1^{\alpha_1} \cdot ... \cdot p_k^{\alpha_k}$ is known, the totient function can be expressed as $\phi(2^n-1) = \prod_{i=1}^k p_i^{\alpha_i-1}(p_i-1)$. Verifying that a polynomial of degree $n$ is primitive also relies on knowing the factorization of $2^n-1$, and these factorizations are used in Ref.~\cite{Zivkovic1994}.
\end{enumerate}
Importantly, inefficient scaling is not an issue in the NISQ era, since both finding primitive polynomials and determining $\phi(2^n-1)$ is possible for all $n<352$. 
Furthermore, a benchmark with efficient scaling can be obtained with a small modification. The classical evaluation would scale efficiently if determining the success ratio was only required for values of $n$ for which the factorization of $2^n-1$ is known. Such $n$ values do exist and are listed up to above 4000 in Ref.~\cite{Zivkovic1994}, for instance.
We expect that once quantum computers reach $n_s \approx 4000$, it will be possible to run Shor's factoring algorithm as a benchmark.

\item 
\emph{Shot number per quantum circuit}:
The shot number $10^4$ in step 3 is chosen to ensure that the statistical error of each empirical success probability is less than $1\%$.

\item 
\emph{Number of qubits in the control register}:
This benchmark leaves freedom to adjust
the size $t$ of the control register used for the phase estimation.
In the original version of Shor's period finding algorithm, $t$ is slightly above $2n$~\cite{Shor1997, nielsen2010quantum}.
Alternatively, the size of the control register can be reduced to just $1$, if mid-circuit measurements and feed-forward are available on the quantum computer~\cite{Beauregard2003}.

\item 
\emph{The smallest meaningful benchmark score}:
The threshold value $0.15$ in Eq.~\eqref{eq:threshold} is somewhat arbitrarily chosen based on the following considerations: A random-number generator that outputs uniformly distributed bitstrings can achieve a success ratio of $\eta \approx 0.18$ for $n = 3$ and $\eta \approx 0.12$ for $n=4$. Therefore, the threshold value $0.15$ already distinguishes a quantum computer and a random-number generator for $n=4$, but not yet for $n=3$. Consequently, the smallest meaningful benchmark score is $n_s = 4$.

\edm

\section{Bell-state quantum error correction benchmark}
\label{sec:qec-benchmark}

Quantum error correction is essential for large-scale quantum computations. 
This benchmark evaluates a quantum computer’s readiness to do quantum computing using quantum error correction. 
Specifically, it measures how much the error (infidelity) of a Bell state can be reduced by preparing it using quantum error correction with respect to preparing it directly on the physical qubits of the same quantum computer. 

\subsection{Background}

Quantum error correction~\cite{Roffe03072019, Lidar_Brun_2013, nielsen2010quantum} is essential for running large-scale quantum computations consisting of deep quantum circuits with negligible error rates.
It entails redundant encoding of quantum information in so-called logical qubits, distributed across several physical qubits
% Janos: This point is already covered in the references before, no need to cite{PhysRevA.52.R2493}
to support the detection and correction of errors during the quantum computation. It requires 
sophisticated circuit design so errors do not propagate in an uncontrolled fashion, and sufficiently small error rates on the components. Currently, quantum computers are not yet at the scale where quantum error correction can be employed on problems of interesting size. However, the first steps towards fault-tolerant quantum computing (quantum computing on logical qubits with quantum error correction) have already been demonstrated~\cite{Paetznick_2024, Bluvstein2023, Bluvstein2025, SalesRodriguez2025,Reichardt2024}. 
% Janos: I suggest to omit references that are about quantum memory, and not FTQC:
% suppressingQuantumErrors, Andersen2020, Andersen_2019, cite-key, Negnevitsky_2018, Ofek_2016, Hu_2019,   PhysRevX.11.041058,

Bell states are maximally entangled states of two qubits whose preparation, error correction, and measurement are computationally relevant tasks. The error-corrected (fault-tolerant) preparation of Bell states has been realized on several current quantum computers~\cite{Paetznick_2024, Reichardt2024, Bluvstein2025,Lacroix2025}. 
The Bell states, 
\begin{align}
\label{eq:Bell_state_stabilizer_def}
\ket{\Phi^\pm} &= \frac{1}{\sqrt{2}} \left( \ket{00} \pm \ket{11}\right);&
\ket{\Psi^\pm} &= \frac{1}{\sqrt{2}} \left( \ket{01} \pm \ket{10}\right), 
\end{align}
form an orthogonal basis on two qubits, and are eigenstates of the operators $\hat{Z}\otimes \hat{Z}$ and $\hat{X}\otimes\hat{X}$, with eigenvalues $\pm 1$ (the 4 combination of eigenvalues correspond to the 4 Bell states). 
A Bell measurement is a projective measurement of the two-qubit state with four outcomes, which correspond to the four Bell states---the measurement of two two-qubit observables $\hat{X}\otimes\hat{X}$ and $\hat{Z}\otimes{Z}$ in a single experiment. 
This can be done, e.g., by running a Bell-state preparation circuit in reverse: applying a CNOT to the two qubits followed by an X measurement of the control qubit and a Z measurement of the target qubit.

The quality of a noisy Bell state can be evaluated by repeated measurements to estimate the Bell fidelity. 
Using repeated Bell measurements on noisy realization $\hat \rho$ of a target Bell state  $\ket{\text{Bell}}$, this reads  
\begin{align}
\label{eq:Bell_infidelity_estimator_direct}
\mathcal{F} = \bra{\text{Bell}} \hat{\rho} \ket{\text{Bell}}
\approx F =  1 - \frac{\text{Number of Bell errors}}{\text{Number of measurements}},
\end{align}
%\begin{align}
%    \mathcal{F}  
%\end{align}
%\{ \ket{\Phi^\pm}, \ket{\Psi^\pm}\}$. Using Bell measurements, %the fidelity can be estimated by 
where the first equation defines the Bell fidelity $\mathcal{F}$, and the last equation defines its estimator $F$, %$\expect{\mathcal{F}}$, 
with a "Bell error" being one of the 3 outcomes of the Bell measurement different from the intended one. If $\ket{\text{Bell}} = \ket{\Phi^+}$, for example, a "Bell error" is if the Bell measurement returns $XX=-1$ and/or $ZZ=-1$.  

The Bell fidelity can also be estimated directly from single-qubit measurements on the two qubits constituting the state, by measuring Pauli correlations~\cite{Paetznick_2024}. 
For this, after preparation of the Bell state, both of its qubits should be measured in the same, randomly chosen basis, $X$, $Y$, or $Z$. 
For any target Bell state, the two outcomes should either be perfectly correlated or anticorrelated (depending on the basis and on the target state). Thus, two of the four outcomes is consistent with the target state, and the other two outcomes correspond to a ``stabilizer error".
The estimator for the Bell fidelity reads
\begin{align}
\label{eq:Bell_infidelity_estimator_stabilizer}
 F &= 1-\frac{3}{2} \frac{\text{Number of stabilizer errors}}{\text{Number of measurements}}.
\end{align}
To be self-contained, we include a short derivation of this result in Sec.~\ref{sec:qec-remarks}.
%Here, to explain a ``stabilizer error", we describe the measurement protocol. The steps are: (i) a Bell state is prepared, (ii) a readout basis ($X$, $Y$, or $Z$) is chosen at random, with equal probability, (iii) both qubits of the state are read out in that basis, 
%One way to realize this is, for $Z$ basis the two qubits should be read out; for $X$, these readouts should be preceded by a Hadamard applied to both qubits independently; for $Y$, the readouts should be preceded by an S gate and a Hadamard applied to both qubits independently. 
%and (iv)  

\subsection{The benchmark}

The benchmark aims to quantify how well a quantum computer is equipped to perform quantum error correction during a quantum computation. 
The benchmarking task is to use quantum error correction on the quantum computer to obtain a high-fidelity Bell state on logical qubits. The benchmark score measures how superior the fidelity of this Bell state is compared to a Bell state prepared on the same hardware on physical qubits.  
%The benchmark score $\mathcal{Q}$ (physical-to-logical ratio of infidelities -- detailed below) provides a comparison between the Bell fidelity  measured on the logical qubits and the highest achievable Bell fidelity measured on the physical qubits on the same hardware. 

For the Bell state obtained using quantum error correction, we require at least one round of active quantum error correction after the preparation and before the measurement; this involves 
quantum operations conditioned on the syndrome (either obtained via mid-circuit measurement or implemented fully coherently without measurement). 
%Additionally, the quantum error correction code employed must have a distance $d$ of at least $3$, meaning the error correction code can correct at least $1$ error.
Non-scalable methods such as post-selection (e.g., on flags) are not permitted. Error suppression, e.g., using dynamical decoupling, is permitted. 

For creating the Bell states, the provider may employ the best quantum circuit on the given hardware and the given quantum error correcting code. Because the four Bell states are equivalent to each other up to local Pauli operations, we allow for protocols that produce one of them at random, if it is indicated which Bell state was produced.

The determination of the Bell fidelity should be done by repeatedly running the preparation circuit and performing suitable measurements on the output, which are either consistent with the Bell state or report an error. 
%The measurements can be direct Bell measurements, or can 
%one of the protocols outlined in Sec.~\ref{sec:ghz-benchmark}, or by 
%consist of single-qubit measurements on the two qubits---we detail two approaches below.

%A Bell measurement is a projective measurement of the two-qubit state with four outcomes, which correspond to the four Bell states. It amounts to the measurement of two two-qubit observables $XX$ and $ZZ$ in a single experiment, e.g., by running a Bell-state preparation circuit in reverse: applying a CNOT to the two qubits followed by an X measurement of the control qubit and a Z measurement of the target qubit. An \emph{error} is one of the 3 outcomes corresponding to the Bell states different from the one intended to be produced by the circuit. 

%Specifically, to obtain $\expect{Z_j Z_k}$ the qubits $j$ and $k$ should be both read out; to obtain $\expect{X_j X_k}$ this readout should be preceded by a Hadamard applied to both qubits independently; to obtain $\expect{Y_j Y_k}$ this readout should be preceded by an S gate and a Hadamard applied to both qubits independently.

\subsection{Step-by-step description of the benchmark protocol}

\begin{enumerate}
\item \textbf{Measure the best Bell fidelity on physical qubits:} \\
Measure the best Bell fidelity achievable on two physical qubits on your hardware, without using quantum error correction or quantum error mitigation. You should choose the pair of physical qubits best suited for the task. Repeat the preferred measurement protocol enough times to gather a precise estimate for the fidelity.  

\item \textbf{Measure the best Bell fidelity on a Bell state of two logical qubits:}\\
Measure the logical-level Bell fidelity, with at least one round of quantum error correction, repeating the following for as many rounds as required for high enough precision:
\begin{enumerate}
    \item Prepare a Bell state of two logical qubits, of at least distance 3. You should know which Bell state it is---if this is decided during the running of the circuit, you should also have this information (logical Pauli frame).   
    \item Perform at least one round of quantum error correction (on two orthogonal bases) on the Bell state, with error correction on the hardware level: either by extracting the stabilizer values by mid-circuit measurements and feed-forward (ideally beyond Pauli-frame updates, to support future fault-tolerant quantum computation), or using quantum gates controlled by the ancilla values to implement the error correction.
    %\sout{a quantum operation conditioned on the results of the decoding (rather than just a classical basis update).}
    \item Measure the two-qubit state, using either a Bell measurement or by measuring one of the stabilizers at random. Record an error if that is the measurement outcome.
\end{enumerate}

\item \textbf{Calculate the bechmark score:}\\
Use the formula given below to evaluate the benchmark score $\mathcal Q$.

\end{enumerate}

\subsection{Definition of the benchmark score}

The benchmark score $\mathcal{Q}$ is defined as the ratio of the measured physical Bell infidelity ($1-$fidelity) to the logical Bell infidelity. In a formula, 
\begin{align}
\mathcal{Q} &= \frac{1 - \text{max}_{jl}F_{jl}}{1-\bar{F}}.
\end{align}
Here, $F$ denotes the measured estimate of the Bell fidelity, obtained by counting errors, as per Eq.~\eqref{eq:Bell_infidelity_estimator_direct} or by Eq.~\eqref{eq:Bell_infidelity_estimator_stabilizer}. We use $F_{jl}$ for the Bell states prepared on physical qubits $j$ and $l$, and $\bar{F}$ for the Bell state prepared on the logical qubits.   

\subsection{Numerical performance analysis}

We performed numerical simulation to explore the Bell-state quantum error correction benchmark for noisy quantum computers realizing lattice surgery on surface code patches~\cite{Horsman2012}. 
We assumed a square grid of qubits, with nearest neighbor connectivity, as in many current superconducting quantum computers\cite{Acharya2025,Kosen2024,gao2024establishingnewbenchmarkquantum}.
We approximated the noise on the qubits and the gates using random Pauli operators, which allowed us to run circuit-level simulations (including noise on the quantum circuit used for syndrome extraction) for large code sizes using Clifford simulation, specifically, STIM~\cite{Gidney_STIM}, with Pymatching~\cite{Higgott_2025}. We varied the surface-code patch sizes $d$, where each patch has $d\times d$ physical data qubits, and $d$ is the code distance. We also varied the noise rates, using parameters for qubit, gate, and measurement noise representative of current quantum computers---using a single physical error rate $p$ to control the error levels on all the physical components~\cite{Gidney2021}, as detailed in Table~\ref{tab:errors}, and also exploring the impact of the two-qubit gate and measurement errors separately.   
We found that a code score $\mathcal{Q}$ above 1 is within reach for a superconducing quantum computer with $\sim 50$ physical qubits (two surface-17 patches), while for $2<\mathcal{Q}<10$ more qubits and larger code patches are required. We detail our findings below.

\begin{table}
    \centering % instead of \begin{center}
\begin{tabular}{c|c|c|c|c|c|c|c|}
%\hline
~ & CNOT & 1Q gate& Init & Readout & Idle & 
Resonator Idle\\
\hline
Standard depolarizing (SD6) & $p$ & $p$ &$p$ &$p$ &$p$ &$p$ \\
\hline
Superconducting Inspired (SI1000) & $p$ & $p/10$ &$2p$ &$5p$ &$p/10$ &$2p$ \\
\hline
\end{tabular}
\caption{\label{tab:errors}
Error levels on the physical components and operations according to two schemes~\cite{Gidney2021}. Here, ``1Q gate" refers to any single-qubit Clifford gate, ``Init" is initialization, and ``Resonator Idle" is the error incurred on idling data qubits while the ancilla qubits are read out. }
\end{table}

For the numerical simulation of the measurement of the Bell infidelity on two physical qubits of a noisy quantum computer, we simulated Pauli correlation experiments. Here the steps were: 
\begin{enumerate}
    \item Initialize the two qubits in $\ket{0}$.
    \item Run a Hadamard on qubit 1, and a CNOT with qubit 1 as the control qubit; this results in the creation of a Bell state $\ket{\Phi^+}$.
    \item[3a.] If the $Z$ values are required, read out the two qubits.
    \item[3b.] If the $X$ values are required, perform a Hadamard on both qubits and then read them out.
    \item[3c.] If the $Y$ values are required, perform a Hadamard and an S gate on both qubits and then read them out.  
\end{enumerate}
For each data point we repeated the above cycle $3$ million times
($10^6$ times for each of the $X$, $Y$, and $Z$ basis).

For the numerical simulation of the measurement of the Bell infidelity on error-corrected qubits, we simulated the creation of a Bell pair on two patches of the surface code using a merge and split, and the estimation of the Bell fidelity by measuring Pauli correlations on the two logical qubits. This was realized by the following protocol: 
\begin{enumerate}
    \item Initialize the data qubits of the two patches in state $\ket{+}$, and then perform one syndrome extraction round. This results in the fault-tolerant initialization of the two patches (two logical qubits) in the $\ket{+} \otimes \ket{+}$ state.
    \item Realize a fault-tolerant measurement of $\bar{Z}\otimes \bar{Z}$ by a ``merge and split" protocol:
    include an extra column ($d$ extra data qubits and $d+1$ extra stabilizers) between the two patches and performing $d$ rounds of stabilizer meausurements (merge). Perform one additional syndrome extraction round, and read out the extra column of data qubits in the X basis (split). This results in the creation of a logical Bell state $\ket{\Phi^+}$ for measurement outcome of +1, and of a $\ket{\Psi^+}$ for a measurement outcome of $-1$.
    \item[3ab.] If the $X$ or $Z$ logical values are required, read out all the data qubits in the $X$ or $Z$ basis, and decode to obtain these; or  
    \item[3c.] If the $Y$ logical values are required, perform perfect syndrome extraction, decode, and perform perfect readout of all data qubits in the $Y$ basis.
\end{enumerate}
For each data point we repeated the above cycle $3$ million times
($10^6$ times per logical Pauli basis). Note that here the measurement of the $Y$ values is not from full simulation of a fault-tolerant protocol: although such protocols have been developed~\cite{Gidney2024} for the surface code, their simulation is more involved (and likely increasing the logical error rate somewhat); we leave this for future work.   
%estimate the error of the fault-tolerant readout in the logical $Y$ basis 
%, we calculated the Bell infidelities with both perfect and noisy logical $X$ and $Z$ measurements (see fig.\ref{fig:qec_scan_p}). 
%We expect that faithfully simulating fault-tolerant logical $Y$ measurements would increase the gap between the "perfect" and "noisy" final readouts by a factor of   

\begin{figure}
    \centering
    \includegraphics[width=0.49\linewidth]{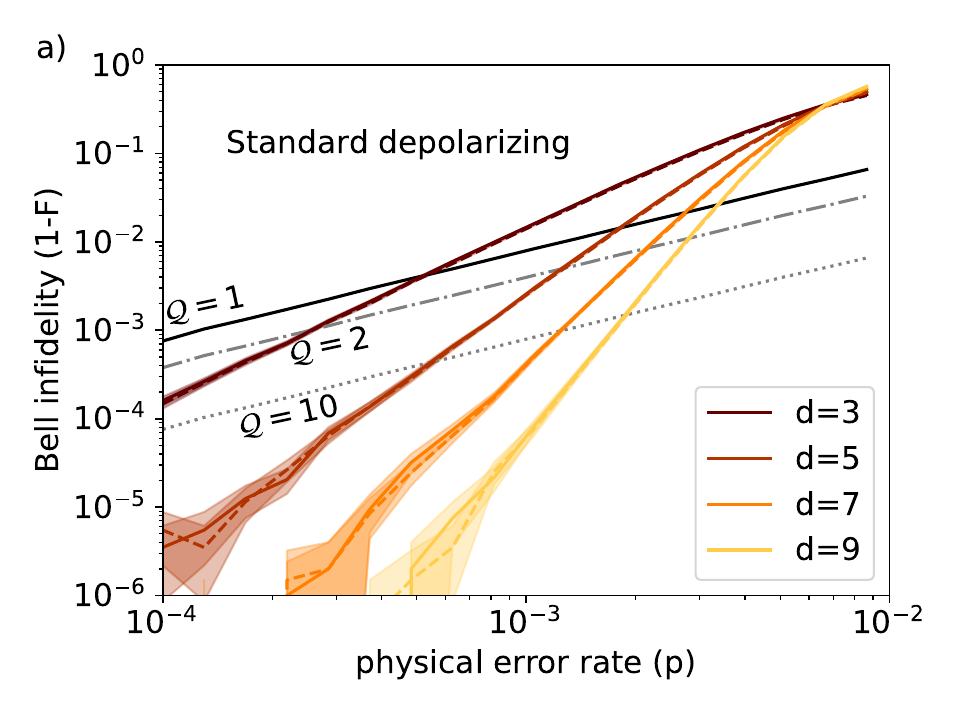}
    \includegraphics[width=0.49\linewidth]{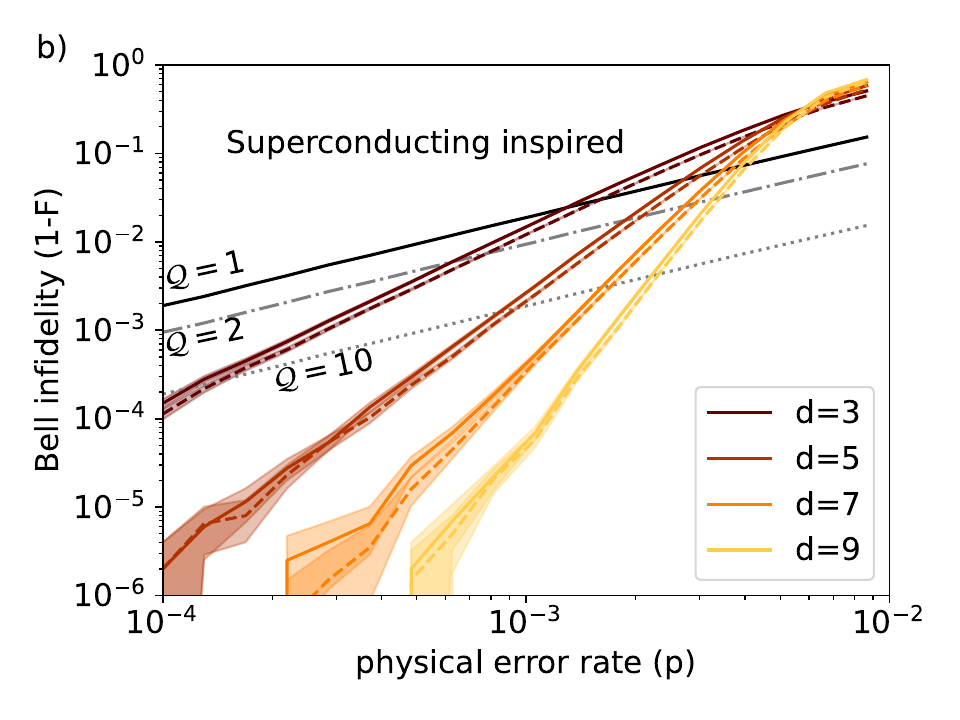}
    \caption{The Bell infidelity as the function of physical error rate for (a) standard depolarizing and (b) superconducting-inspired noise models; see Table~\ref{tab:errors}. Bell infidelities from surface-code lattice surgery protocols with noisy $X$, $Z$ and perfect $Y$ final measurements for $d = 3, 5, 7, 9$ are shown as colored continuous lines,  with shaded regions corresponding to $\pm2\sigma$ error bars. To help assess the possible impact of noisy $Y$ measurements, results with all final measurements perfect are also shown with dashed lines and shaded regions. Simulations on physical circuits on 2 qubits are shown as black and gray lines, with infidelities divided by $\mathcal{Q}$ as indicated---these serve as $\mathcal{Q}$-isolines. Each data point shown is calculated from $3\cdot 10^6$ runs.} 
    \label{fig:qec_scan_p}
\end{figure}

We explored what Bell infidelity values and $\mathcal Q$ scores can be expected depending on the physical error rates $p$ for various surface-code patch sizes. 
Here we used a single physical error rate $p$ to control the error levels on all the physical components, as detailed in Table~\ref{tab:errors}.  
Our results for 18 different error rates and surface-code patch sizes of $d=3, 5, 7, 9$ are shown in Fig.~\ref{fig:qec_scan_p}. Here the measured Bell infidelity values are shown both for the simulations of the error-corrected lattice surgery protocols (in different colors), and also for the protocols on the physical qubits (black lines). For the extraction of the $\mathcal Q$ scores corresponding to various patch sizes, we included the physical-qubit values divided by $\mathcal{Q}=2$ and $10$, in dash-dotted and dotted lines, respectively; these serve as $\mathcal Q$-isolines, whose intersections with the colored lines correspond to expected $\mathcal{Q}$ scores of 1, 2, 10. As expected, decreasing the error rate $p$ increases the advantage gained by error correction: for $p$ below the threshold, the error-corrected Bell infidelities are proportional to $p^{(d+1)/2}$, while the physical Bell infidelities are proportional to $p$ (fits not shown). The simulations show that for the current quantum computers, which have $p \approx \qty{0.1}{\percent}$, in the standard depolarizing error model, surface-code patches of $d\ge5$ are required for $\mathcal Q >1$, while for the superconducting-inspired error model, already $d=3$ surface-code patches give $\mathcal Q \approx 2$. 
We remark that because of the efficiency of STIM~\cite{Gidney_STIM} and Pymatching~\cite{Higgott_2025}, all these simulations run in Python took approximately 100 minutes of wall time on a MacBook Pro (Apple M3 Pro chip). 

\begin{figure}
    \centering
    \includegraphics[width=0.6\linewidth]{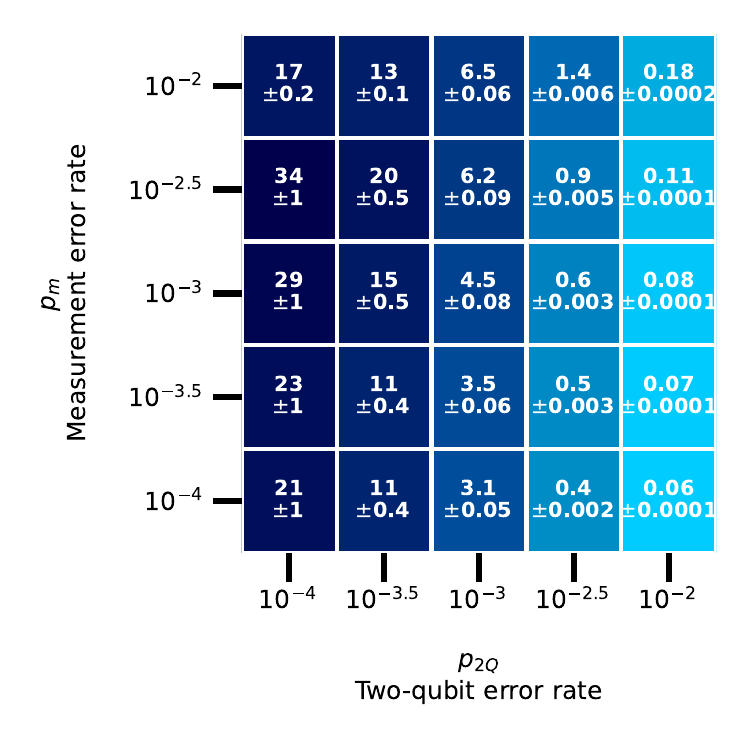}
    \caption{Numerically calculated benchmark scores for the Bell-state quantum error correction benchmark for different measurement and two-qubit-gate error rates. The logical Bell infidelities correspond to logical qubits encoded into distance-5 surface-code patches. The other error rates were fixed: 1Q gate: $10^{-4}$; Init: $2\cdot10^{-3}$; Idle: $10^{-4}$; Resonator Idle: $2\cdot 10^{-3}$.}
    \label{fig:qec_qscore}
\end{figure}

We further explored how varying the rate of two-qubit errors and of measurement errors independently impacts the benchmark score $Q$. For this we took a patch size of $d=5$, and used as baseline the superconducting-inspired error model with $p=\qty{0.1}{\percent}$. We then chose 25 different combinations of measurement error rate and CNOT error rate, and evaluated numerically the Bell infidelities for the physical qubits and for the logical qubits, as detailed above. We show the numerically calculated $\mathcal Q$ scores in Fig.~\ref{fig:qec_qscore}. We observe that the $\mathcal{Q}$ score is above 1 if the two-qubit gates are good enough (error rate is at or below $10^{-3}$). Interestingly, the $\mathcal{Q}$ score is usually higher for higher measurement rates when the rate of two-qubit errors is fixed. This is a result of the physical circuit being more sensitive to measurement noise than the logical counterpart as long as the measurement errors are not too strong compared to other noise sources.

%For both numerical estimators given above for the Bell infidelity,  it is straightforward to estimate their precision, since the number of errors should follow a Poisson distribution. 

\subsection{Remarks}
\label{sec:qec-remarks}

\begin{enumerate}
\item \emph{Variation: insist on mid-circuit measurements?}
In the current version, we allow for measurement-free quantum error correction, where no mid-circuit measurements are performed. This approach can be better than doing mid-circuit measurements on current hardware. However, it could be unfavourable on larger quantum computers as it is not expected to be scalable.  
\item \emph{Variation: more strict on Bell measurement.} We could require a true Bell-state measurement that would extract 2 bits from the state, and would be in the ideal case a projective measurement on the four Bell states.  
\item \emph{Generalizations: connection to the other benchmarks.} 
The Bell states (all four are equivalent up to local Pauli operators) can be seen as the smallest GHZ state, so this benchmark is closely related to the GHZ benchmark in Sec.~\ref{sec:ghz-benchmark}, and uses some of its specifications.  
In the same way, quantum error-corrected versions of the other benchmarks could also be of interest -- even though, for the time being, it is unclear when the period finding or Clifford Volume can be implemented with logical qubits. 
%We envision that the same benchmarks can, however, be employed to demonstrate the benefit of quantum error correction and logical qubits:
This generalization would have several advantages:
(a) the Clifford Volume would highlight that error rates on logical qubits are ideally lower than on physical qubits, and would therefore offer a larger algorithmic depth. Along the same lines, (b) period finding with logical qubits should yield a higher success rate than with physical qubits. In both cases, the ratio of the error rates may serve as a benchmark score to highlight the benefit of using logical qubits.
\item \emph{Derivation of numerical estimator of the Bell fidelity. }
We briefly derive the numerical estimator for the Bell fidelity given in Eq.~ \eqref{eq:Bell_infidelity_estimator_stabilizer}, if the target Bell state is $\ket{\Phi^\pm}$; rewriting the derivation for $\ket{\Psi^\pm}$ is straightforward.
For this, we write the projectors to the Bell states $\ket{\Phi^\pm}$ as
\begin{align}
\label{eq:Bell_state_stabilizer_def}
\ket{\Phi^\pm}\!\bra{\Phi^\pm} &= \frac{1 \pm \hat{X} \!\otimes\! \hat{X} 
\mp \hat{Y} \!\otimes\! \hat{Y} + \hat{Z} \!\otimes\! \hat{Z}}4.&
%\ket{\Psi^\pm}\!\bra{\Psi^\pm} &= \frac{1 \pm \hat{X} \!\otimes\! \hat{X} 
%\pm \hat{Y} \!\otimes\! \hat{Y} - \hat{Z} \!\otimes\! \hat{Z}}4.
\end{align}
It follows that the Bell fidelity reads,  
\begin{align}
    \mathcal{F}^{\pm} = \bra{\Phi^\pm} \hat{\rho} \ket{\Phi^\pm} =
    \frac{1 \pm\expect{\hat{X}\otimes \hat{X}} 
    \mp\expect{\hat{Y} \otimes \hat{Y}}
    +\expect{\hat{Z} \otimes \hat{Z}}}4.
\end{align}
%with similar formulas for the states $\ket{\Psi^\pm}$. 
These expectation values can be obtained by repeatedly preparing the state and measuring the two qubits constituting the state in the same basis. Specifically, 
%taking the Bell state $\ket{\Phi^\pm}$, and 
using the notation $N_Z$ for the number of times the $Z$ basis was chosen, and $N_{ZZ=-1}$ to denote the number of times an error was found, i.e., unequal values of the two qubits were measured, with analogous notation for the $X$ and $Y$ basis, we have
\begin{align}
1-{F^\pm} = \frac{1}{2} \left( \frac{N_{ZZ=-1}}{N_Z} + 
\frac{N_{XX=\mp 1}}{N_X} + \frac{N_{YY=\pm1}}{N_Y} \right) 
\approx \frac{3}{2} \frac{\text{Number of errors}}{\text{Number of measurements}},
\end{align}
%where $\approx$ here denotes the unbiased statistical estimator,
where we used $1- {2 N_{ZZ=-1}}/{N_Z}$ as unbiased estimator for $\expect{ZZ}$, with analogous estimators for $\expect{XX}$ and $\expect{YY}$, and for the last approximate equality  
took $N_X \approx N_Y \approx N_Z$. The right-hand-side here corresponds to Eq.~\eqref{eq:Bell_infidelity_estimator_stabilizer}. 
\end{enumerate}

\subsubsection*{}

% Need to clarify whether we want to restrict or clarify mid-circuit measurement or not vs fully coherent error correction

\section{Discussion and outlook}

In this work, we have presented four core key performance indicators (KPIs) designed to provide a comprehensive assessment of quantum computing capabilities across the late-NISQ and early fault-tolerant eras. Each benchmark addresses a distinct yet complementary aspect of quantum processor performance.  An important characteristic of this benchmarking suite is its commitment to reproducibility and transparency. In addition to providing detailed protocol specifications, we have made available open-source implementations and standardized evaluation routines. This approach ensures that the benchmarks can be consistently executed across different platforms and that results can be fairly compared. The availability of reference implementations also lowers barriers to adoption and enables continuous refinement of the benchmarking methodology through community feedback.

We emphasize that the benchmarks presented here are not static endpoints but rather starting points for an ongoing dialogue within the quantum computing community.  We anticipate and encourage modifications initiated by community feedback, which will be essential for ensuring that the benchmarking suite remains relevant and useful as the field progresses. The open-source nature of our implementation facilitates such evolution by enabling collaborative refinement and extension of the protocols.

%The above overview of the core KPIs outlines four directions that need to be followed as we scale up quantum computers: from maintaining high-fidelity control and large-scale entanglement, via demonstration of applications, to improvements on removing undesired entropy during the computation. The number of benchmarks to be considered is expected to grow, including aspects such as data rate or power consumption. These numbers cover the needs of specialists, but the community may also need to consider an aggregation of benchmarks that support consolidated communication of progress.

%The goal here is to maintain the possibility to communicate progress in the community. As the field is continuously evolving, we foresee that these benchmarks shall undergo a similar evolution. We also anticipate that modifications will be initiated based on the feedback from the community.

\begin{acknowledgments}

This work was mainly supported by the Horizon Europe programme HORIZON-CL4-2022-QUANTUM-01-SGA via the project 101113946 OpenSuperQPlus100 and by the European Union’s Horizon Europe research and innovation programme under Grant Agreement Number 101114305 via the project MILLENION, as well as by the European Union’s Horizon Europe research and innovation programme under grant agreement No.~101135699 via the project SPINUS.

\end{acknowledgments}

%\section*{Appendix: Technical description of the benchmarks} 
%\section*{Technical description of the Key Performance Indicators}
%\appendix

%\section{Template benchmark structure?}
%\subsection{Background}
%\subsection{The benchmark}
%\subsection{Step-by-step description of computing the benchmark}
%\subsection{Definition of the benchmark score}
%\subsection{Remarks}

%\newpage

%\input{sections/clifford-volume-benchmark}
%\newpage
%\input{sections/ghz-benchmark-v2}
%\newpage
%\input{sections/shor-benchmark}
%\newpage
%\input{sections/qec-benchmark-v2}

%\bibliographystyle{apssamp}
\bibliography{bibliography}

@article{Flammia_2011,
   title={{Direct Fidelity Estimation from Few Pauli Measurements}},
   volume={106},
   pages = {230501},
   DOI={10.1103/physrevlett.106.230501},
   journal={Physical Review Letters},
   author={Flammia, Steven T. and Liu, Yi-Kai},
   year={2011}
}

@article{LiuYangYang,
author = {Liu, Xia and Yang, Huan and Yang, Li},
title = {{Feasibility Analysis of Cracking RSA with Improved Quantum Circuits of the Shor’s Algorithm}},
journal = {Security and Communication Networks},
volume = {2023},
pages = {2963110},
doi = {https://doi.org/10.1155/2023/2963110},
year = {2023}
}

@article{Beauregard2003,
  author       = {St{\'{e}}phane Beauregard},
  title        = {{Circuit for Shor's algorithm using 2n+3 qubits}},
  journal      = {Quantum Information and Computation},
  volume       = {3},
  pages        = {175--185},
  year         = {2003},
  doi          = {10.26421/QIC3.2-8}
}

@article{Hashim2025TutorialQCVV,
  title        = {{A Practical Introduction to Benchmarking and Characterization of Quantum Computers}},
  author       = {Akel Hashim and Long B. Nguyen and Noah Goss and Brian Marinelli and Ravi K. Naik and Trevor Chistolini and Jordan Hines and J. P. Marceaux and Yosep Kim and Pranav Gokhale and Teague Tomesh and Senrui Chen and Liang Jiang and Samuele Ferracin and Kenneth Rudinger and Timothy Proctor and Kevin C. Young and Robin Blume-Kohout and Irfan Siddiqi},
  journal      = {PRX Quantum},
  volume       = {6},
  pages        = {030202},
  year         = {2025},
  doi          = {10.1103/PRXQuantum.6.030202}
}

@misc{lall2025review,
  title={A review and collection of metrics and benchmarks for quantum computers: definitions, methodologies and software},
  author={Lall, Deep and Agarwal, Abhishek and Zhang, Weixi and Lindoy, Lachlan and Lindstr{\"o}m, Tobias and Webster, Stephanie and Hall, Simon and Chancellor, Nicholas and Wallden, Petros and Garcia-Patron, Raul and Elham Kashefi and Viv Kendon and Jonathan Pritchard and Alessandro Rossi and Animesh Datta and Theodoros Kapourniotis and Konstantinos Georgopoulos and Ivan Rungger},
  eprint={2502.06717},
  archivePrefix={arXiv},
  year={2025}
}

@article{PhysRevA.100.032328,
  title = {Validating quantum computers using randomized model circuits},
  author = {Cross, Andrew W. and Bishop, Lev S. and Sheldon, Sarah and Nation, Paul D. and Gambetta, Jay M.},
  journal = {Physical Review A},
  volume = {100},
  issue = {3},
  pages = {032328},
  numpages = {11},
  year = {2019},
  month = {Sep},
  publisher = {American Physical Society},
  doi = {10.1103/PhysRevA.100.032328},
  url = {https://link.aps.org/doi/10.1103/PhysRevA.100.032328}
}

@article{Proctor2021,
  doi = {10.1038/s41567-021-01409-7},
  url = {https://doi.org/10.1038/s41567-021-01409-7},
  year = {2021},
  month = dec,
  publisher = {Springer Science and Business Media {LLC}},
  volume = {18},
  number = {1},
  pages = {75--79},
  author = {Timothy Proctor and Kenneth Rudinger and Kevin Young and Erik Nielsen and Robin Blume-Kohout},
  title = {Measuring the capabilities of quantum computers},
  journal = {Nature Physics}
}

@misc{lorenz2025systematic,
  title={Systematic benchmarking of quantum computers: status and recommendations},
  author={Lorenz, Jeanette Miriam and Monz, Thomas and Eisert, Jens and Reitzner, Daniel and Schopfer, F{\'e}licien and Barbaresco, Fr{\'e}d{\'e}ric and Kurowski, Krzysztof and van der Schoot, Ward and Strohm, Thomas and Senellart, Jean and Cécile M. Perrault and Martin Knufinke and Ziyad Amodjee and Mattia Giardini},
  eprint={2503.04905},
  archivePrefix={arXiv},
  year={2025}
}

@article{Terhal2015,
  title = {Quantum error correction for quantum memories},
  volume = {87},
  ISSN = {1539-0756},
  url = {http://dx.doi.org/10.1103/RevModPhys.87.307},
  DOI = {10.1103/revmodphys.87.307},
  number = {2},
  journal = {Reviews of Modern Physics},
  publisher = {American Physical Society (APS)},
  author = {Terhal,  Barbara M.},
  year = {2015},
  month = apr,
  pages = {307–346}
}

@article{Proctor_2022,
  title = {{Scalable Randomized Benchmarking of Quantum Computers Using Mirror Circuits}},
  author = {Proctor, Timothy and Seritan, Stefan and Rudinger, Kenneth and Nielsen, Erik and Blume-Kohout, Robin and Young, Kevin},
  journal = {Physical Review Letters},
  volume = {129},
  pages = {150502},
  year = {2022},
  doi = {10.1103/PhysRevLett.129.150502}
}

@article{eisert2020quantum,
  title={Quantum certification and benchmarking},
  author={Eisert, Jens and Hangleiter, Dominik and Walk, Nathan and Roth, Ingo and Markham, Damian and Parekh, Rhea and Chabaud, Ulysse and Kashefi, Elham},
  journal={Nature Reviews Physics},
  volume={2},
  number={7},
  pages={382--390},
  year={2020},
  doi = {10.1038/s42254-020-0186-4},
  publisher={Nature Publishing Group UK London}
}

@INPROCEEDINGS{ShorAlgorithm,
  author={Shor, P.W.},
  booktitle={Proceedings 35th Annual Symposium on Foundations of Computer Science}, 
  title={Algorithms for quantum computation: discrete logarithms and factoring}, 
  year={1994},
  volume={},
  number={},
  pages={124-134},
  doi={10.1109/SFCS.1994.365700}
}

@article{Entanglement,
  author = {J. Richard and L. Noah},
  doi = {10.1098/rspa.2002.1097},
  journal = {Proceedings of the Royal Society A},
  pages = {2011--2032},
  title = {On the role of entanglement in quantum-computational speed-up},
  volume = {459},
  year = {2003},
}

@article{PhysRevLett.91.147902,
  author = {G. Vidal},
  doi = {https://journals.aps.org/prl/abstract/10.1103/PhysRevLett.91.147902},
  journal = {Physical Review Letters},
  pages = {147902},
  title = {Efficient Classical Simulation of Slightly Entangled Quantum Computations},
  volume = {91},
  year = {2003},
}

@article{Verstraete2008,
  author = {F. Verstraete and V. Murg and J. I. Cirac},
  doi = {10.1080/14789940801912366},
  journal = {Advances in Physics},
  number = {2},
  pages = {143--224},
  title = {Matrix product states, projected entangled pair states, and variational renormalization group methods for quantum spin systems},
  volume = {57},
  year = {2008},
}

@article{RevModPhys.90.035005,
  author = {L. Pezzè and A. Smerzi and M. K. Oberthaler and R. Schmied and P. Treutlein},
  doi = {https://link.aps.org/doi/10.1103/RevModPhys.90.035005},
  journal = {Reviews of Modern Physics},
  pages = {035005},
  title = {Quantum metrology with nonclassical states of atomic ensembles},
  volume = {90},
  year = {2018},
}

@article{QuantumTeleportation,
  author = {P. Espoukeh and P. Pedram},
  doi = {doi:10.1007/s11128-014-0766-2},
  journal = {Quantum Information Processing},
  pages = {1789-1811},
  title = {Quantum teleportation through noisy channels with multi-qubit {GHZ} states},
  volume = {13},
  year = {2014},
}

@article{Man2006,
  author = {Z.-X. Man and Y.-J. Xia and B. A. Nguyen},
  journal = {Journal of Physics B: Atomic, Molecular and Optical Physics},
  doi = {10.1088/0953-4075/39/18/015},
  pages = {3855},
  title = {Quantum secure direct communication by using {GHZ} states and entanglement swapping},
  volume = {39},
  year = {2006},
}

@incollection{Greenberger1989,
  author = {D. M. Greenberger and M. A. Horne and A. Zeilinger},
  booktitle = {Bell's Theorem, Quantum Theory and Conceptions of the Universe},
  doi = {10.1007/978-94-017-0849-4\_10},
  editor = {Menas Kafatos},
  pages = {69--72},
  publisher = {Springer Dordrecht},
  series = {Fundamental Theories of Physics},
  title = {Going Beyond {B}ell’s Theorem},
  volume = {37},
  year = {1989},
}

@article{Song2019,
  author = {Chao Song  and Kai Xu  and Hekang Li  and Yu-Ran Zhang  and Xu Zhang  and Wuxin Liu  and Qiujiang Guo  and Zhen Wang  and Wenhui Ren  and Jie Hao  and Hui Feng  and Heng Fan  and Dongning Zheng  and Da-Wei Wang  and H. Wang  and Shi-Yao Zhu},
  doi = {https://www.science.org/doi/abs/10.1126/science.aay0600},
  journal = {Science},
  number = {6453},
  pages = {574-577},
  title = {Generation of multicomponent atomic {S}chrödinger cat states of up to 20 qubits},
  volume = {365},
  year = {2019},
}

@article{Omran2019,
  author = {A. Omran  and H. Levine  and A. Keesling  and G. Semeghini  and T. T. Wang  and S. Ebadi  and H. Bernien  and A. S. Zibrov  and H. Pichler  and S. Choi  and J. Cui  and M. Rossignolo  and P. Rembold  and S. Montangero  and T. Calarco  and M. Endres  and M. Greiner  and V. Vuleti\'{c}  and M. D. Lukin},
  doi = {https://www.science.org/doi/abs/10.1126/science.aax9743},
  journal = {Science},
  number = {6453},
  pages = {570-574},
  title = {Generation and manipulation of {S}chrödinger cat states in {R}ydberg atom arrays},
  volume = {365},
  year = {2019},
}

@article{Zhong2018,
  author = {Zhong, Han-Sen and Li, Yuan and Li, Wei and Peng, Li-Chao and Su, Zu-En and Hu, Yi and He, Yu-Ming and Ding, Xing and Zhang, Weijun and Li, Hao and Zhang, Lu and Wang, Zhen and You, Lixing and Wang, Xi-Lin and Jiang, Xiao and Li, Li and Chen, Yu-Ao and Liu, Nai-Le and Lu, Chao-Yang and Pan, Jian-Wei},
  doi = {https://link.aps.org/doi/10.1103/PhysRevLett.121.250505},
  journal = {Physical Review Letters},
  pages = {250505},
  title = {12-Photon Entanglement and Scalable Scattershot Boson Sampling with Optimal Entangled-Photon Pairs from Parametric Down-Conversion},
  volume = {121},
  year = {2018},
}

@article{Guhne2010,
  author = {O. Gühne and M. Seevinck},
  doi = {10.1088/1367-2630/12/5/053002},
  journal = {New Journal of Physics},
  pages = {053002},
  title = {Separability criteria for genuine multiparticle entanglement},
  volume = {12},
  year = {2010},
}

@article{Cruz2019,
  author = {Cruz, Diogo and Fournier, Romain and Gremion, Fabien and Jeannerot, Alix and Komagata, Kenichi and Tosic, Tara and Thiesbrummel, Jarla and Chan, Chun Lam and Macris, Nicolas and Dupertuis, Marc-André and Javerzac-Galy, Clément},
  doi = {10.1002/qute.201900015},
  journal = {Advanced Quantum Technologies},
  number = {5-6},
  pages = {1900015},
  title = {Efficient Quantum Algorithms for {GHZ} and {W} States, and Implementation on the {IBM} Quantum Computer},
  volume = {2},
  year = {2019},
}

@article{Ozaeta2019,
  author = {A. Ozaeta and P. L. McMahon},
  doi = {10.1088/2058-9565/ab13e5},
  journal = {Quantum Science and Technology},
  pages = {025015},
  title = {Decoherence of up to 8-qubit entangled states in a 16-qubit superconducting quantum processor},
  volume = {4},
  year = {2019},
}

@article{Blatt2008,
  author = {R. Blatt and D. Wineland},
  doi = {doi:10.1038/nature07125},
  journal = {Nature},
  pages = {1008-1015},
  title = {Entangled States of Trapped Atomic Ions},
  volume = {453},
  year = {2008},
}

@article{Moses2023,
  author = {Moses, S. A. and others},
  doi = {10.1103/PhysRevX.13.041052},
  journal = {Physical Review X},
  pages = {041052},
  title = {A Race-Track Trapped-Ion Quantum Processor},
  volume = {13},
  year = {2023},
}

@article{Mooney2021,
  author = {Gary J. Mooney and Gregory A. L. White and Charles D. Hill and Lloyd C. L. Hollenberg},
  doi = {10.1088/2399-6528/ac1df7},
  journal = {Journal of Physics Communications},
  pages = {095004},
  title = {Generation and verification of 27-qubit {G}reenberger--{H}orne--{Z}eilinger states in a superconducting quantum computer},
  volume = {5},
  year = {2021},
}

@article{Leibfried2005,
  author = {Dietrich Leibfried and others},
  doi = {10.1038/nature04251},
  journal = {Nature},
  pages = {639-642},
  title = {Creation of a six-atom ‘{S}chrödinger cat’ state},
  volume = {438},
  year = {2005},
}

@article{PhysRevA.101.032343,
  author = {Wei, Ken X. and Lauer, Isaac and Srinivasan, Srikanth and Sundaresan, Neereja and McClure, Douglas T. and Toyli, David and McKay, David C. and Gambetta, Jay M. and Sheldon, Sarah},
  doi = {10.1103/PhysRevA.101.032343},
  journal = {Physical Review A},
  title = {{Verifying multipartite entangled Greenberger-Horne-Zeilinger states via multiple quantum coherences}},
  pages = {032343},
  volume = {101},
  year = {2020},
}

@article{PhysRevLett.119.180511,
  author = {Song, Chao and Xu, Kai and Liu, Wuxin and Yang, Chui-ping and Zheng, Shi-Biao and Deng, Hui and Xie, Qiwei and Huang, Keqiang and Guo, Qiujiang and Zhang, Libo and Zhang, Pengfei and Xu, Da and Zheng, Dongning and Zhu, Xiaobo and Wang, H. and Chen, Y.-A. and Lu, C.-Y. and Han, Siyuan and Pan, Jian-Wei},
  doi = {https://link.aps.org/doi/10.1103/PhysRevLett.119.180511},
  journal = {Physical Review Letters},
  pages = {180511},
  title = {10-Qubit Entanglement and Parallel Logic Operations with a Superconducting Circuit},
  volume = {119},
  year = {2017},
}

@article{Barends2014SuperconductingQC,
  author = {Rami Barends and others},
  doi = {10.1038/nature13171},
  journal = {Nature},
  pages = {500-503},
  title = {Superconducting quantum circuits at the surface code threshold for fault tolerance},
  volume = {508},
  year = {2014},
}

@article{Huang2020,
  author = {Hsin-Yuan Huang and Richard Kueng and John Preskill},
  doi = {10.1038/s41567-020-0932-7},
  journal = {Nature Physics},
  pages = {1050--1057},
  title = {Predicting Many Properties of a Quantum System from Very Few Measurements},
  volume = {16},
  year = {2020},
}

@unpublished{David,
  archiveprefix = {arXiv},
  author = {A. Acuaviva and D. Aguirre and R. Peña and M. Sanz},
  doi = {doi:10.48550/arXiv.2407.10941},
  eprint = {2407.10941},
  title = {Benchmarking Quantum Computers: Towards a Standard Performance Evaluation Approach},
  year = {2024},
}

@article{Knill2005,
  author = {E. Knill},
  doi = {doi:10.1038/nature03350},
  journal = {Nature},
  pages = {39-44},
  title = {Quantum computing with realistically noisy devices},
  volume = {434},
  year = {2005},
}

@article{Takeda2022,
  author = {Kenta Takeda and Akito Noiri and Takashi Nakajima and Takashi Kobayashi and Seigo Tarucha},
  doi = {doi:10.1038/s41586-022-04986-6},
  journal = {Nature},
  pages = {682-686},
  title = {Quantum error correction with silicon spin qubits},
  volume = {608},
  year = {2022},
}

@article{Graham2022,
  author = {T. M. Graham and others},
  doi = {10.1038/s41586-022-04603-6},
  journal = {Nature},
  pages = {457-462},
  title = {Multi-qubit entanglement and algorithms on a neutral-atom quantum computer},
  volume = {604},
  year = {2022},
}

@article{PhysRevLett.117.210502,
  author = {Wang, Xi-Lin and Chen, Luo-Kan and Li, W. and Huang, H.-L. and Liu, C. and Chen, C. and Luo, Y.-H. and Su, Z.-E. and Wu, D. and Li, Z.-D. and Lu, H. and Hu, Y. and Jiang, X. and Peng, C.-Z. and Li, L. and Liu, N.-L. and Chen, Yu-Ao and Lu, Chao-Yang and Pan, Jian-Wei},
  doi = {10.1103/PhysRevLett.117.210502},
  journal = {Physical Review Letters},
  pages = {210502},
  title = {Experimental Ten-Photon Entanglement},
  volume = {117},
  year = {2016},
}

@article{huang2024certifyingquantumstatessinglequbit,
    url = {https://www.nature.com/articles/s41567-025-03025-1},
    author = {Hsin-Yuan Huang and John Preskill and Mehdi Soleimanifar},
    title = {Certifying almost all quantum states with few single-qubit measurements},
    journal = {Nature Physics},
    volume = {21},
    pages = {1834–1841},
    year = {2025}
}

@article{kam2023generationpreservationlargeentangled,
  author = {John F. Kam and Haiyue Kang and Charles D Hill and Gary J Mooney and Lloyd C L Hollenberg},
  journal = {Physical Review Research},
  title = {Characterization of entanglement on superconducting quantum computers of up to 414 qubits},
  doi = {10.1103/PhysRevResearch.6.033155},
volume = {6},
pages = {033155},
  year = {2024},
}

@article{1,
  author = {Aaronson, Scott and Gottesman, Daniel},
  title = {Improved simulation of stabilizer circuits},
  year = {2004},
  journal = {Physical Review A},
  volume = {70},
  pages = {052328},
  doi = {10.1103/physreva.70.052328},
  url = {http://dx.doi.org/10.1103/physreva.70.052328},
}

@article{4,
  author = {Bravyi, Sergey and Maslov, Dmitri},
  title = {Hadamard-Free Circuits Expose the Structure of the {C}lifford Group},
  year = {2021},
  journal = {IEEE Transactions on Information Theory},
  volume = {67},
  pages = {4546-4563},
  number = {7},
  doi = {10.1109/tit.2021.3081415},
  url = {http://dx.doi.org/10.1109/tit.2021.3081415},
}

@unpublished{7,
  archiveprefix = {arXiv},
  author = {Matthew DeCross and Reza Haghshenas and Minzhao Liu and Enrico Rinaldi and Johnnie Gray and Yuri Alexeev and Charles H. Baldwin and John P. Bartolotta and Matthew Bohn and Eli Chertkov and Julia Cline and Jonhas Colina and Davide DelVento and Joan M. Dreiling and Cameron Foltz and John P. Gaebler and Thomas M. Gatterman and Christopher N. Gilbreth and Joshua Giles and Dan Gresh and Alex Hall and Aaron Hankin and Azure Hansen and Nathan Hewitt and Ian Hoffman and Craig Holliman and Ross B. Hutson and Trent Jacobs and Jacob Johansen and Patricia J. Lee and Elliot Lehman and Dominic Lucchetti and Danylo Lykov and Ivaylo S. Madjarov and Brian Mathewson and Karl Mayer and Michael Mills and Pradeep Niroula and Juan M. Pino and Conrad Roman and Michael Schecter and Peter E. Siegfried and Bruce G. Tiemann and Curtis Volin and James Walker and Ruslan Shaydulin and Marco Pistoia and Steven. A. Moses and David Hayes and Brian Neyenhuis and Russell P. Stutz and Michael Foss-Feig},
  eprint = {arXiv:2406.02501},
  title = {The computational power of random quantum circuits in arbitrary geometries},
  year = {2024},
}

@article{Gidney2021,
  title = {{A Fault-Tolerant Honeycomb Memory}},
  volume = {5},
  DOI = {10.22331/q-2021-12-20-605},
  journal = {Quantum},
  author = {Gidney,  Craig and Newman,  Michael and Fowler,  Austin and Broughton,  Michael},
  year = {2021},
  pages = {605}
}

@article{Gidney2024,
  title = {{Inplace Access to the Surface Code Y Basis}},
  volume = {8},
  DOI = {10.22331/q-2024-04-08-1310},
  journal = {Quantum},
  author = {Gidney,  Craig},
  year = {2024},
  pages = {1310}
}

@article{Horsman2012,
  title = {Surface code quantum computing by lattice surgery},
  volume = {14},
  DOI = {10.1088/1367-2630/14/12/123011},
  journal = {New Journal of Physics},
  author = {Horsman,  Dominic and Fowler,  Austin G and Devitt,  Simon and Meter,  Rodney Van},
  year = {2012},
  pages = {123011}
}

@article{Gidney_STIM,
  author = {Gidney, Craig},
  title = {Stim: a fast stabilizer circuit simulator},
  year = {2021},
  journal = {Quantum},
  volume = {5},
  pages = {497},
  doi = {10.22331/q-2021-07-06-497},
  url = {http://dx.doi.org/10.22331/q-2021-07-06-497},
}

@article{Higgott_2025,
   title={{Sparse Blossom: correcting a million errors per core second with minimum-weight matching}},
   volume={9},
   DOI={10.22331/q-2025-01-20-1600},
   journal={Quantum},
   author={Higgott, Oscar and Gidney, Craig},
   year={2025},
   pages={1600}
}

@article{PhysRevA.99.042337,
  title = {Validating and certifying stabilizer states},
  author = {Kalev, Amir and Kyrillidis, Anastasios and Linke, Norbert M.},
  journal = {Physical Review A},
  volume = {99},
  issue = {4},
  pages = {042337},
  numpages = {6},
  year = {2019},
  month = {Apr},
  publisher = {American Physical Society},
  doi = {10.1103/PhysRevA.99.042337},
  url = {https://link.aps.org/doi/10.1103/PhysRevA.99.042337}
}

@article{proctor2024benchmarking,
  title   = {Benchmarking quantum computers},
  author  = {Proctor, Timothy and Young, Kevin and Baczewski, Andrew D. and Blume-Kohout, Robin},
  journal = {Nature Reviews Physics},
  volume  = {7},
  pages   = {105--118},
  year    = {2025},
  doi     = {10.1038/s42254-024-00796-z}
}

@article{Shor1997,
  title = {Polynomial-Time Algorithms for Prime Factorization and Discrete Logarithms on a Quantum Computer},
  volume = {26},
  ISSN = {1095-7111},
  url = {http://dx.doi.org/10.1137/S0097539795293172},
  DOI = {10.1137/s0097539795293172},
  number = {5},
  journal = {SIAM Journal on Computing},
  publisher = {Society for Industrial & Applied Mathematics (SIAM)},
  author = {Shor,  Peter W.},
  year = {1997},
  month = oct,
  pages = {1484-1509}
}

@article{Greenberger1990,
  title = {Bell’s theorem without inequalities},
  volume = {58},
  ISSN = {1943-2909},
  url = {http://dx.doi.org/10.1119/1.16243},
  DOI = {10.1119/1.16243},
  number = {12},
  journal = {American Journal of Physics},
  publisher = {American Association of Physics Teachers (AAPT)},
  author = {Greenberger,  Daniel M. and Horne,  Michael A. and Shimony,  Abner and Zeilinger,  Anton},
  year = {1990},
  month = dec,
  pages = {1131–1143}
}

@article{Monz2011,
  title = {14-Qubit Entanglement: Creation and Coherence},
  volume = {106},
  ISSN = {1079-7114},
  url = {http://dx.doi.org/10.1103/PhysRevLett.106.130506},
  DOI = {10.1103/physrevlett.106.130506},
  number = {13},
  journal = {Physical Review Letters},
  publisher = {American Physical Society (APS)},
  author = {Monz,  Thomas and Schindler,  Philipp and Barreiro,  Julio T. and Chwalla,  Michael and Nigg,  Daniel and Coish,  William A. and Harlander,  Maximilian and H\"{a}nsel,  Wolfgang and Hennrich,  Markus and Blatt,  Rainer},
  year = {2011},
  pages = {130506}
}

@article{Pogorelov2021,
  title = {Compact Ion-Trap Quantum Computing Demonstrator},
  volume = {2},
  ISSN = {2691-3399},
  url = {http://dx.doi.org/10.1103/PRXQuantum.2.020343},
  DOI = {10.1103/prxquantum.2.020343},
  number = {2},
  journal = {PRX Quantum},
  publisher = {American Physical Society (APS)},
  author = {Pogorelov,  I. and Feldker,  T. and Marciniak,  Ch. D. and Postler,  L. and Jacob,  G. and Krieglsteiner,  O. and Podlesnic,  V. and Meth,  M. and Negnevitsky,  V. and Stadler,  M. and H\"{o}fer,  B. and W\"{a}chter,  C. and Lakhmanskiy,  K. and Blatt,  R. and Schindler,  P. and Monz,  T.},
  year = {2021},
  pages = {020343} 
}

@article{Bluvstein2023,
  title = {Logical quantum processor based on reconfigurable atom arrays},
  volume = {626},
  ISSN = {1476-4687},
  url = {http://dx.doi.org/10.1038/s41586-023-06927-3},
  DOI = {10.1038/s41586-023-06927-3},
  number = {7997},
  journal = {Nature},
  author = {Bluvstein,  Dolev and Evered,  Simon J. and Geim,  Alexandra A. and Li,  Sophie H. and Zhou,  Hengyun and Manovitz,  Tom and Ebadi,  Sepehr and Cain,  Madelyn and Kalinowski,  Marcin and Hangleiter,  Dominik and Bonilla Ataides,  J. Pablo and Maskara,  Nishad and Cong,  Iris and Gao,  Xun and Sales Rodriguez,  Pedro and Karolyshyn,  Thomas and Semeghini,  Giulia and Gullans,  Michael J. and Greiner,  Markus and Vuletić,  Vladan and Lukin,  Mikhail D.},
  year = {2024},
  pages = {58–65}
}

@article{Bluvstein2025,
  title = {A fault-tolerant neutral-atom architecture for universal quantum computation},
  ISSN = {1476-4687},
  url = {http://dx.doi.org/10.1038/s41586-025-09848-5},
  DOI = {10.1038/s41586-025-09848-5},
  journal = {Nature},
  publisher = {Springer Science and Business Media LLC},
  author = {Bluvstein,  Dolev and Geim,  Alexandra A. and Li,  Sophie H. and Evered,  Simon J. and Bonilla Ataides,  J. Pablo and Baranes,  Gefen and Gu,  Andi and Manovitz,  Tom and Xu,  Muqing and Kalinowski,  Marcin and Majidy,  Shayan and Kokail,  Christian and Maskara,  Nishad and Trapp,  Elias C. and Stewart,  Luke M. and Hollerith,  Simon and Zhou,  Hengyun and Gullans,  Michael J. and Yelin,  Susanne F. and Greiner,  Markus and Vuletić,  Vladan and Cain,  Madelyn and Lukin,  Mikhail D.},
  year = {2025},
  month = nov 
}

@article{SalesRodriguez2025,
  title = {Experimental demonstration of logical magic state distillation},
  volume = {645},
  ISSN = {1476-4687},
  url = {http://dx.doi.org/10.1038/s41586-025-09367-3},
  DOI = {10.1038/s41586-025-09367-3},
  number = {8081},
  journal = {Nature},
  publisher = {Springer Science and Business Media LLC},
  author = {Sales Rodriguez and others},
  year = {2025},
  month = jul,
  pages = {620–625}
}

@misc{Reichardt2024,
  eprint = {2409.04628},
  author = {Reichardt,  Ben W. and Aasen,  David and Chao,  Rui and Chernoguzov,  Alex and van Dam,  Wim and Gaebler,  John P. and Gresh,  Dan and Lucchetti,  Dominic and Mills,  Michael and Moses,  Steven A. and Neyenhuis,  Brian and Paetznick,  Adam and Paz,  Andres and Siegfried,  Peter E. and da Silva,  Marcus P. and Svore,  Krysta M. and Wang,  Zhenghan and Zanner,  Matt},
  title = {Demonstration of quantum computation and error correction with a tesseract code},
  archivePrefix = {arXiv},
  year = {2024}
}

@article{Lacroix2025,
  title = {Scaling and logic in the colour code on a superconducting quantum processor},
  volume = {645},
  ISSN = {1476-4687},
  url = {http://dx.doi.org/10.1038/s41586-025-09061-4},
  DOI = {10.1038/s41586-025-09061-4},
  number = {8081},
  journal = {Nature},
  publisher = {Springer Science and Business Media LLC},
  author = {Lacroix,  N. and others},
  year = {2025},
  month = may,
  pages = {614–619}
}

@misc{Paetznick_2024,
  eprint = {2404.02280},
  author = {Paetznick,  A. and others},
  title = {Demonstration of logical qubits and repeated error correction with better-than-physical error rates},
  archivePrefix = {arXiv},
  year = {2024}
}

@misc{coppersmith2002approximate,
  title={An approximate {F}ourier transform useful in quantum factoring},
  author={Coppersmith, Don},
  eprint={quant-ph/0201067},
  archivePrefix = {arXiv},
  year={2002}
}

@book{nielsen2010quantum,
  title={Quantum computation and quantum information},
  author={Nielsen, Michael A and Chuang, Isaac L},
  year={2010},
  publisher={Cambridge University Press}
}

@article{Chen2024benchmarkingtrapped,
  doi = {10.22331/q-2024-11-07-1516},
  url = {https://doi.org/10.22331/q-2024-11-07-1516},
  title = {Benchmarking a trapped-ion quantum computer with 30 qubits},
  author = {Chen, Jwo-Sy and Nielsen, Erik and Ebert, Matthew and Inlek, Volkan and Wright, Kenneth and Chaplin, Vandiver and Maksymov, Andrii and P{\'{a}}ez, Eduardo and Poudel, Amrit and Maunz, Peter and Gamble, John},
  journal = {Quantum},
  issn = {2521-327X},
  publisher = {Verein zur F{\"{o}}rderung des Open Access Publizierens in den Quantenwissenschaften},
  volume = {8},
  pages = {1516},
  year = {2024}
}

@misc{IQMs,
      title={Technology and Performance Benchmarks of {IQM}'s 20-Qubit Quantum Computer}, 
    author={Leonid Abdurakhimov and others},
      year={2024},
      eprint={2408.12433},
      archivePrefix={arXiv}
}

@article{Henriet2020quantumcomputing,
  doi = {10.22331/q-2020-09-21-327},
  url = {https://doi.org/10.22331/q-2020-09-21-327},
  title = {Quantum computing with neutral atoms},
  author = {Henriet, Lo{\"{i}}c and Beguin, Lucas and Signoles, Adrien and Lahaye, Thierry and Browaeys, Antoine and Reymond, Georges-Olivier and Jurczak, Christophe},
  journal = {Quantum},
  issn = {2521-327X},
  publisher = {Verein zur F{\"{o}}rderung des Open Access Publizierens in den Quantenwissenschaften},
  volume = {4},
  pages = {327},
  month = sep,
  year = {2020}
}

@article{Photonic,
author = {Rad, H. and Ainsworth, T. and Alexander, R. and Altieri, B. and Askarani, M. and Baby, Reenu and Banchi, L. and Baragiola, Ben and Bourassa, J. and Chadwick, R. and Charania, I. and Chen, H. and Collins, M. and Contu, P. and D’Arcy, N. and Dauphinais, Guillaume and De Prins, Robbe and Deschenes, D. and Luch, I. and Zhang, Y.},
year = {2025},
pages = {912--919},
title = {Scaling and networking a modular photonic quantum computer},
volume = {638},
journal = {Nature},
doi = {10.1038/s41586-024-08406-9}
}

@article{Chen_2023,
   title={Linear cross-entropy benchmarking with {C}lifford circuits},
   volume={108},
   DOI={10.1103/physreva.108.052613},
   number={5},
   journal={Physical Review A},
   pages={052613},
   author={Chen, Jianxin and Ding, Dawei and Huang, Cupjin and Kong, Linghang},
   year={2023}}

@misc{Merkel2025CliffordProxy,
  title        = {When {C}lifford Benchmarks Are Sufficient: Estimating Application Performance with Scalable Proxy Circuits},
  author       = {Susan A. Merkel and Timothy Proctor and Samuele Ferracin and Jordan Hines and Samantha Barron and Luke C. G. Govia and David McKay},
  year         = {2025},
  eprint       = {2503.05943},
  archivePrefix = {arXiv}
}

@book{Lidar_Brun_2013, place={Cambridge}, title={Quantum Error Correction}, 
author={Daniel A Lidar and Todd A Brun},
publisher={Cambridge University Press}, year={2013}}

@article{Roffe03072019,
author = {Joschka Roffe},
title = {Quantum error correction: an introductory guide},
journal = {Contemporary Physics},
volume = {60},
number = {3},
pages = {226--245},
year = {2019},
publisher = {Taylor \& Francis},
doi = {10.1080/00107514.2019.1667078}
}

@article{Kosen2024,
  title = {Signal Crosstalk in a Flip-Chip Quantum Processor},
  volume = {5},
  ISSN = {2691-3399},
  url = {http://dx.doi.org/10.1103/PRXQuantum.5.030350},
  DOI = {10.1103/prxquantum.5.030350},
  number = {3},
  journal = {PRX Quantum},
  publisher = {American Physical Society (APS)},
  author = {Kosen,  Sandoko and Li,  Hang-Xi and Rommel,  Marcus and Rehammar,  Robert and Caputo,  Marco and Gr\"{o}nberg,  Leif and Fernández-Pendás,  Jorge and Kockum,  Anton Frisk and Biznárová,  Janka and Chen,  Liangyu and Križan,  Christian and Nylander,  Andreas and Osman,  Amr and Roudsari,  Anita Fadavi and Shiri,  Daryoush and Tancredi,  Giovanna and Govenius,  Joonas and Bylander,  Jonas},
  year = {2024},
  month = sep 
}

@misc{gao2024establishingnewbenchmarkquantum,
      title={Establishing a New Benchmark in Quantum Computational Advantage with 105-qubit Zuchongzhi 3.0 Processor}, 
      author={Dongxin Gao and others},
      year={2024},
      eprint={2412.11924},
      archivePrefix={arXiv},
      primaryClass={quant-ph},
      url={https://arxiv.org/abs/2412.11924}, 
}

@article{Zivkovic1994,
  author = {Miodrag Živković},
  title = {A Table of Primitive Binary Polynomials},
  journal = {Mathematics of Computation},
  volume = {62},
  number = {205},
  pages = {385--386},
  year = {1994},
  doi = {10.1090/S0025-5718-1994-1201073-4}
}

@article{Patel2008,
    author = {Patel, Ketan N. and Markov, Igor L. and Hayes, John P.},
    title = {Optimal synthesis of linear reversible circuits},
    year = {2008},
    issue_date = {March 2008},
    publisher = {Rinton Press, Incorporated},
    address = {Paramus, NJ},
    volume = {8},
    number = {3},
    issn = {1533-7146},
    journal = {Quantum Information \& Computation},
    month = mar,
    pages = {282–294},
    url = {https://dl.acm.org/doi/10.5555/2011763.2011767}
}

@misc{davis2021benchmarksquantumcomputersshors,
      title={{Benchmarks for quantum computers from Shor's algorithm}}, 
      author={E. D. Davis},
      year={2021},
      eprint={2111.13856},
      archivePrefix={arXiv}
}

@Article{Genting2023,
    AUTHOR = {Dai, Genting and He, Kaiyong and Zhao, Changhao and He, Yongcheng and Liu, Jianshe and Chen, Wei},
    TITLE = {{Quasi-Shor Algorithms for Global Benchmarking of Universal Quantum Processors}},
    JOURNAL = {Applied Sciences},
    VOLUME = {13},
    YEAR = {2023},
    pages = {139},
    DOI = {10.3390/app13010139}
}

@misc{gidney2025factor2048bitrsa,
      title={{How to factor 2048 bit RSA integers with less than a million noisy qubits}}, 
      author={Craig Gidney},
      year={2025},
      eprint={2505.15917},
      archivePrefix={arXiv}
}

@article{Gidney2021howtofactorbit,
  doi = {10.22331/q-2021-04-15-433},
  url = {https://doi.org/10.22331/q-2021-04-15-433},
  title = {How to factor 2048 bit {RSA} integers in 8 hours using 20 million noisy qubits},
  author = {Gidney, Craig and Eker{\aa{}}, Martin},
  journal = {Quantum},
  issn = {2521-327X},
  publisher = {Verein zur F{\"{o}}rderung des Open Access Publizierens in den Quantenwissenschaften},
  volume = {5},
  pages = {433},
  month = apr,
  year = {2021}
}

@ARTICLE{Lubinski2023,
author={Lubinski, Thomas and Johri, Sonika and Varosy, Paul and Coleman, Jeremiah and Zhao, Luning and Necaise, Jason and Baldwin, Charles H. and Mayer, Karl and Proctor, Timothy},
journal={IEEE Transactions on Quantum Engineering},
title={{Application-Oriented Performance Benchmarks for Quantum Computing}},
year={2023},
volume={4},
pages={1-32},
url = {https://doi.ieeecomputersociety.org/10.1109/TQE.2023.3253761},
}

@unpublished{CliffordVolume2025,
  author = {Portik, Attila and K{\'a}lm{\'a}n, Orsolya and Monz, Thomas and Zimbor{\'a}s, Zolt{\'a}n},
  title = {Clifford Volume and Free Fermion Volume: Complementary Scalable Benchmarks for Quantum Computers},
  note = {Manuscript in preparation},
  year = {2025}
}

@article{Acharya2025,
author = {Acharya, Rajeev and others},
doi = {10.1038/s41586-024-08449-y},
journal = {Nature},
pages = {920--926},
title = {{Quantum error correction below the surface code threshold}},
volume = {638},
year = {2025}
}

@article{Kim2023,
author = {Kim, Youngseok and Eddins, Andrew and Anand, Sajant and Wei, Ken Xuan and van den Berg, Ewout and Rosenblatt, Sami and Nayfeh, Hasan and Wu, Yantao and Zaletel, Michael and Temme, Kristan and Kandala, Abhinav},
doi = {10.1038/s41586-023-06096-3},
journal = {Nature},
pages = {500},
title = {{Evidence for the utility of quantum computing before fault tolerance}},
volume = {618},
year = {2023}
}

@article{Gao2025,
author = {Gao, Dongxin and others},
doi = {10.1103/PhysRevLett.134.090601},
journal = {Physical Review Letters},
pages = {090601},
title = {{Establishing a New Benchmark in Quantum Computational Advantage with 105-qubit Zuchongzhi 3.0 Processor}},
volume = {134},
year = {2025}
}

@misc{Ransford2025,
archivePrefix = {arXiv},
author = {Ransford, Anthony and others},
eprint = {2511.05465},
title = {{Helios: A 98-qubit trapped-ion quantum computer}},
year = {2025}
}

@article{Edlbauer2025,
author = {Edlbauer, Hermann and Wang, Junliang and Huq, A. M. Saffat-Ee and Thorvaldson, Ian and Jones, Michael T. and Misha, Saiful Haque and Pappas, William J. and Moehle, Christian M. and Hsueh, Yu-Ling and Bornemann, Henric and Gorman, Samuel K. and Chung, Yousun and Keizer, Joris G. and Kranz, Ludwik and Simmons, Michelle Y.},
doi = {10.1038/s41586-025-09827-w},
journal = {Nature},
pages = {569--575},
title = {{An 11-qubit atom processor in silicon}},
volume = {648},
year = {2025}
}

@article{Horodecki2009,
author = {Horodecki, R. and Horodecki, P. and Horodecki, M. and Horodecki, K.},
doi = {10.1103/RevModPhys.81.865},
journal = {Reviews of Modern Physics},
pages = {865},
title = {{Quantum entanglement}},
volume = {81},
year = {2009}
}

@misc{Javadi-Abhari2024,
archivePrefix = {arXiv},
arxivId = {2405.08810},
author = {Javadi-Abhari, Ali and Treinish, Matthew and Krsulich, Kevin and Wood, Christopher J. and Lishman, Jake and Gacon, Julien and Martiel, Simon and Nation, Paul D. and Bishop, Lev S. and Cross, Andrew W. and Johnson, Blake R. and Gambetta, Jay M.},
eprint = {2405.08810},
title = {{Quantum computing with Qiskit}},
year = {2024}
}

@online{EQCB,
  title  = {European Quantum Computing Benchmarks},
  year   = {2025},
  howpublished = {https://gitlab.com/qcpi/eqcb}}

\end{document}